\documentclass[a4paper,11pt]{article}
\pdfoutput=1
\usepackage{jheppub}
\usepackage{color}
\usepackage{xcolor}
\usepackage{array}
\newcolumntype{C}[1]{>{\centering\arraybackslash}p{#1}}
\usepackage{slashed,dsfont}
\usepackage{bm,wasysym}
\usepackage{longtable}
\usepackage{booktabs}
\usepackage{slashed}
\usepackage{tikz}
\usetikzlibrary{positioning,arrows}
\usetikzlibrary{decorations.pathmorphing}
\usetikzlibrary{decorations.markings}

\usepackage{subfigure}

\usepackage{tabularx}
\usepackage{rotating}
\usepackage{lscape} 
\usepackage{hyperref}
\usepackage[normalem]{ulem}
\usepackage{braket}
\newcommand{\lsim}{
\mathrel{\hbox{\rlap{\hbox{\lower4pt\hbox{$\sim$}}}\hbox{$<$}}}}

\newcommand{\gsim}{
\mathrel{\hbox{\rlap{\hbox{\lower4pt\hbox{$\sim$}}}\hbox{$>$}}}}

\renewcommand{\arraystretch}{2}

\def\tr{\text{tr}}

\allowdisplaybreaks[2]

\newcommand{\nn}{\nonumber}

\def\re{{\rm Re}}  \def\im{{\rm Im}}

\def\mathB#1{{\mathcal{B}_{{#1}}}}
\def\mB#1{{m_{\mathcal{B}_{{#1}}}}} 
\def\mmB#1{{m^2_{\mathcal{B}_{{#1}}}}} 

\definecolor{schrift}{RGB}{120,0,0}

\title{\boldmath\color{schrift}{Sterile neutrinos in $\Lambda_b^0\to (\Lambda_c^+,p^+)\ell^-_1\ell^-_2\ell^+_3\nu$ decays}}

\author[a]{Diganta Das}
\author[b]{Jaydeb Das} 
\affiliation[a]{{\sf Center for Computational Natural Sciences and Bioinformatics, International Institute of Information Technology, Hyderabad 500 032, India}}
\affiliation[b]{{\sf Department of Physics and Astrophysics, University of Delhi, Delhi 110007, India}}
\emailAdd{diganta.das@iiit.ac.in}
\emailAdd{jaydebphysics@gmail.com}

\abstract{We study lepton number violating and lepton number conserving semi-leptonic decays of heavy baryons $\Lambda_b^0$ to three charged leptons and a neutrino. The decays occur through two intermediate quasi-degenerate GeV-scale sterile neutrinos of either Majorana or Dirac type that can be on-shell. Interference between the intermediate heavy neutrinos leads to CP violation in the final states. Effect of neutrino oscillations between the heavy states are also considered in observables of interests, \emph{i.e.,} branching ratio, and the CP-asymmetry. Given the present constraints on the heavy-to-light mixing elements $|V_{eN}|$ and $|V_{\mu N}|$, CP-averaged branching ratio of $\Lambda_b^0\to (\Lambda_c,p)\mu\mu e\nu$ with intermediate Majorana neutrinos is almost two orders of magnitude larger than for the same with Dirac neutrinos, whereas, CP-averaged branching ratio of $\Lambda_b^0\to (\Lambda_c,p)ee\mu\nu$ is of the same order of magnitude for both Majorana and Dirac neutrino induced decays. CP-violation is found to be appreciable when the neutrino mass difference is comparable with the average decay widths.} 

\keywords{Baryonic Decays, Sterile neutrino, Majorana neutrino, Lepton number violation}

\begin{document}

\tikzset{
fermion/.style={solid,draw=black, postaction={decorate},decoration={markings,mark=at position 0.8 with {\arrow[draw=black,thick]{>}}}},
antifermion/.style={solid,draw=black, postaction={decorate},decoration={markings,mark=at position 0.8 with {\arrow[draw=black,thick]{<}}}},
majoranafermion/.style={solid,draw=black, postaction={decorate},decoration={markings,mark=at position 0.5 with {\arrow[draw=black,thick]{><}}}},
vector/.style={decorate, decoration={snake, amplitude=0.8mm, segment length=2mm, post length=0.1mm}, draw}
}

\maketitle

\renewcommand{\arraystretch}{1.6}

\section{Introduction}
It is now confirmed from neutrino oscillation experiments that active light neutrinos have mass \cite{Fukuda:1998mi,Wendell:2010md,Ambrosio:2003yz}, albeit very tiny. The reason for the smallness is yet unknown, but the very existence of neutrino mass may indicate the existence of a right-handed gauge-singlet (sterile) counterpart $N_R$. Whether the $N_R$ is of the Dirac or Majorana type is a crucial question and is intricately related to the question of the origin of neutrino mass itself. If $N_R$ is a Dirac type then, tree-level Dirac neutrino mass terms are allowed, but it leads to an unnaturally small Yukawa coupling for neutrinos. If the $N_R$ is a Majorana, then both Dirac type mass term $m_D(\overline{\nu}_L N_R + {\rm h.c})$ and Majorana term $m_N N_R N_R$ are allowed by the SM. Then, small light neutrino mass $m_{\nu }\sim m_D^2/m_N$, where $m_D$ is at the electroweak scale or lower, arise through ``seesaw'' mechanism \cite{Mohapatra:1979ia,Schechter:1980gr,Schechter:1981cv,Minkowski:1977sc,Yanagida:1979as,Ramond:1979py,Levy:1980ws}. There are several versions of the mechanism. In type-I seesaw, the masses if $N_R$ are of the order of a few TeV which leads to $m_\nu\lesssim 1$eV. Recently, low energy seesaw mechanism has been proposed where $N_R$ have a mass of the order few hundreds of MeV to a few GeV \cite{Buchmuller:1991ce,Asaka:2005an,delAguila:2007ap,He:2009ua,Kersten:2007vk,Ibarra:2010xw,Nemevsek:2012cd, Pilaftsis:1991ug}. These GeV-scale seesaw mechanisms can be experimentally accessed at the intensity and the energy frontier. 

Majorana neutrinos distinguish themselves from Dirac through $|\Delta L| =2$ lepton number violating (LNV) processes. Neutrino-less double beta decay ($0\nu\beta\beta$) process is highly sensitive probe of LNV process for light Majorana exchange \cite{Pas:2015eia,Rodejohann:2011mu,DellOro:2016tmg,GomezCadenas:2011it} but enhancement due to GeV-scale Majorana has recently been shown \cite{Drewes:2016lqo, Asaka:2016zib}. Due to the lack of any $0\nu\beta\beta$ processes so far, and due to its limitations to probe LNV in the first neutrino family only, alternative processes are highly desirable. Majorana neutrino-mediated rare decays of mesons and baryons are useful in this regard. Interestingly, if Majorana neutrinos exist in nature, then it will lead to both LNV as well as lepton-number-conserving (LNC) decays of hadrons.  The LNC decays are not unique to Majorana neutrinos and occur through Dirac neutrino exchange as well. The aim of this paper is to study LNV and LNC in $\Lambda_b^0$ baryonic decays. Usually, for a too light or a too heavy sterile neutrino, decay rates are too suppressed to be observed in the current experiments. But, if the sterile neutrino masses are within a few hundreds of MeV to a few GeV, then they can go on-shell resulting in appreciable decay rates \cite{Cvetic:2010rw,Helo:2010cw}. A number of searches of Majorana neutrino mediated LNV decays are underway at the LHC and Belle-II which has renewed its interests in the theoretical aspects including in hadronic decays \cite{Das:2021prm,Helo:2010cw, Cvetic:2010rw, Atre:2005eb, Dib:2000wm, Ali:2001gsa, Zhang:2010um, Yuan:2013yba, Godbole:2020doo, Cvetic:2020lyh, Shuve:2016muy, Chun:2019nwi, Mandal:2017tab, Abada:2017jjx, Abada:2019bac, Mejia-Guisao:2017nzx, Zhang:2021wjj, Cvetic:2019shl, Barbero:2013fc, Mandal:2016hpr, Zamora-Saa:2016qlk,Cvetic:2015ura,Cvetic:2015naa,Cvetic:2014nla,Cvetic:2013eza,Kim:2018uht,Kim:2019xqj,Milanes:2018aku,Mejia-Guisao:2017gqp,Milanes:2016rzr,Castro:2013jsn,Quintero:2011yh,Littenberg:1991rd, Cvetic:2017vwl, Barbero:2002wm}, $\tau$-lepton decays \cite{Castro:2012gi,Dib:2011hc,Yuan:2017xdp,Zamora-Saa:2016ito,Kim:2017pra, Abada:2021zcm}, and different scattering processes \cite{Das:2017zjc,Das:2017rsu,Das:2017nvm,Cvetic:2019rms,Cvetic:2018elt,Fuks:2020zbm,Fuks:2020att,Cai:2017mow,Ruiz:2020cjx,Najafi:2020dkp, Bray:2007ru}. Experimentally, the $B^-\to\pi^+\mu^-\mu^-$ decay has been searched by the LHCb \cite{Aaij:2014aba} and the $K^-\to\pi^+\mu^-\mu^-$ decay has been searched by the NA48/2 Collaborations \cite{CERNNA48/2:2016tdo} and these measurements give the stringent constraints on the heavy-to-light mixing elements. With large integrated luminosity expected at the future Belle-II as well as at the LHC in the future upgrade, the sensitivity of sterile neutrinos to hadronic decays is expected to increase. 

In this paper we study the rare decay processes $\Lambda_b^0\to (\Lambda_c^+,p^+) \ell_1^- \ell_2^- \ell_3^+ \nu $ and their conjugate modes where $\ell_1, \ell_2, \ell_3$ can in general be of different flavors. The decays are induced by sterile neutrinos that can be either of Majorana or Dirac types. Two sterile neutrinos $N_j (j=1,2)$ with masses $m_{N_1}$ and $m_{N_2}$ in the range of a few hundreds of MeV to a few GeV are considered. In this mass ranges, the neutrinos can go on-shell, \emph{i.e.,} decay widths $\Gamma_{N_j}\ll m_{N_j}$. We are interested in scenarios where the two neutrinos are almost degenerate \emph{i.e.,} $\Delta M_N\equiv m_{N_1}-m_{N_2}\ll m_{N_1},m_{N_2}$. Models with quasi-degenerate Majorana neutrinos with mass in this range have been proposed in Ref.~\cite{Dib:2014pga}.  If the sterile neutrinos are Majorana then both LNV and LNC final states will result. If the sterile neutrinos are Dirac, only LNC final state will result. Irrespective of the nature of the neutrinos involved, CP violation is expected for both types of neutrinos. In this paper, we present the expression for the branching ratios for both the LNV and the LNC cases. We note that LNV decays of hyperons are of experimental interest and have been searched by several experimental collaborations in $\Xi, \Lambda_c$ and $\Sigma$ decays \cite{Rajaram:2005bs,Kodama:1995ia,Ablikim:2020xqk}. We are not aware of such searches in $\Lambda_b^0$ decays. Theoretically, only Majorana neutrino induced LNV decays in $\Lambda_b^0\to (\Lambda_c^+, p^+)\pi^+\mu^-\mu^-$ decays were studied in Refs. \cite{Das:2021prm, Mejia-Guisao:2017nzx} where neutrinos oscillation effects were neglected. In this paper, we also include the effect of neutrinos oscillation in the expressions of the branching ratios.

We present CP-averaged branching ratios for $\Delta M_N \sim \Gamma_N$, where $\Gamma_N=(\Gamma_1+\Gamma_2)/2$ is the average decay width. Due to CKM suppression, $\Lambda_b^0 \to p  \ell_1 \ell_2 \ell_3\nu$ has a smaller branching ratio than $\Lambda_b^0 \to \Lambda_c  \ell_1 \ell_2 \ell_3\nu$. When the decays occur through intermediate Majorana neutrinos, branching ratio of  $\Lambda_b ^0\to (\Lambda_c,p)  e e \mu \nu$ is at least two orders of magnitude suppressed compared to $\Lambda_b^0 \to (\Lambda_c,p)\mu\mu e \nu$ due to present experimental upper bounds on the heavy to light mixing elements $|V_{\mu N}|^2 <10^{-5}$, $|V_{e N}|^2 <10^{-7}$. On the other hand, when the decays occur through intermediate Dirac neutrinos, the CP average branching ratios of $\Lambda_b^0 \to (\Lambda_c,p) \mu \mu e \nu$ and $\Lambda_b^0 \to (\Lambda_c,p)  e e \mu \nu$ are of the same order of magnitude. Comparing Majorana and Dirac neutrino induced decays, we find that for $\Lambda_b^0 \to (\Lambda_c,p)  \mu \mu  e \nu$ the branching ratio with intermediate Majorana is almost two orders of magnitude larger than that with intermediate Dirac. But for the $\Lambda_b^0 \to (\Lambda_c,p)  e e \mu \nu$ the branching ratios are of the same order of magnitude for Majorana and Dirac cases. The CP-asymmetry is sensitive to the heavy neutrino displaced vertex length $L$, and maximum CP-violation is obtained in the region where the CP-odd phase is $\pi/2$, $\Delta M_N\sim \Gamma_N$, and $L\sim L_{\rm osc}$, where $L_{\rm osc}$ is the heavy neutrino oscillation length. For large values of $L$, the CP-violation is suppressed.

The paper is organized as follows. In section \ref{fomalism} we give a generic formalism including neutrino oscillation for the case of $\mathcal{B}_1 \to \mathcal{B}_2 \ell_1^- \ell_2^- \ell_3^+\nu $ and its CP conjugate mode $\bar{\mathcal{B}}_1 \to \bar{\mathcal{B}}_2 \ell_1^+ \ell_2^+ \ell_3^-\bar{\nu} $ mediated by degenerate on-shell both Majorana and Dirac neutrinos. In section \ref{Numerical} we perform the numerical analysis and the results are summarized in section \ref{sec:summary}. Some details of our derivations are given in the appendixes.

\section{Formalism \label{fomalism}} 
Our calculations are based on a model where the left-handed neutrinos of the SM $SU(2)$ doublets are accompanied by two right-handed sterile neutrinos $N_1$ and $N_2$. The relation between the SM flavor eigenstates $\nu_{\ell_i L}$ and the mass eigenstates $\nu_{i,L}, N_1, N_2$ is
\begin{equation}
\nu_{\ell_i L} = \sum_{i=1}^3 V_{\ell_i \nu_j}\nu_{j,L} + V_{\ell_i N_1} N_1+ V_{\ell_i N_2} N_2,\,\,\ \ell_i= e^-, \mu^-, \tau^-.
\end{equation}
The heavy to light mixing elements $V_{\ell_i N_1}$ and $V_{\ell_i N_2}$ are the free parameters of our model which can be constrained by experimental data. The parameters in general can be complex. We assume a convention that $V_{\ell_i N_j}$ is the mixing between a negatively charged lepton and a sterile neutrino and define a CP-odd phase $\phi_{\ell_i N_j}$ as
\begin{equation}\label{eq:convention}
V_{\ell_i N_j} = |V_{\ell_i N_j}|e^{i\phi_{\ell_i N_j}}\, ,\quad j=1,2
\end{equation}
Under the assumption of CPT conservation, a Majorana neutrino differs from a Dirac neutrino in the fact that the former needs only two degrees of freedom (two helicities) compared to the latter's four (two helicities each for particle and anti-particle). Another important difference between the two lies in how the PMNS mixing matrix is parametrized for them. For Dirac neutrinos, the PMNS matrix can be parametrized in terms of three mixing angles and one CP-violating phase which is called the Dirac phase \cite{Giunti:2007ry}. For Majorana neutrinos, two additional CP phases, called Majorana phases, are needed.

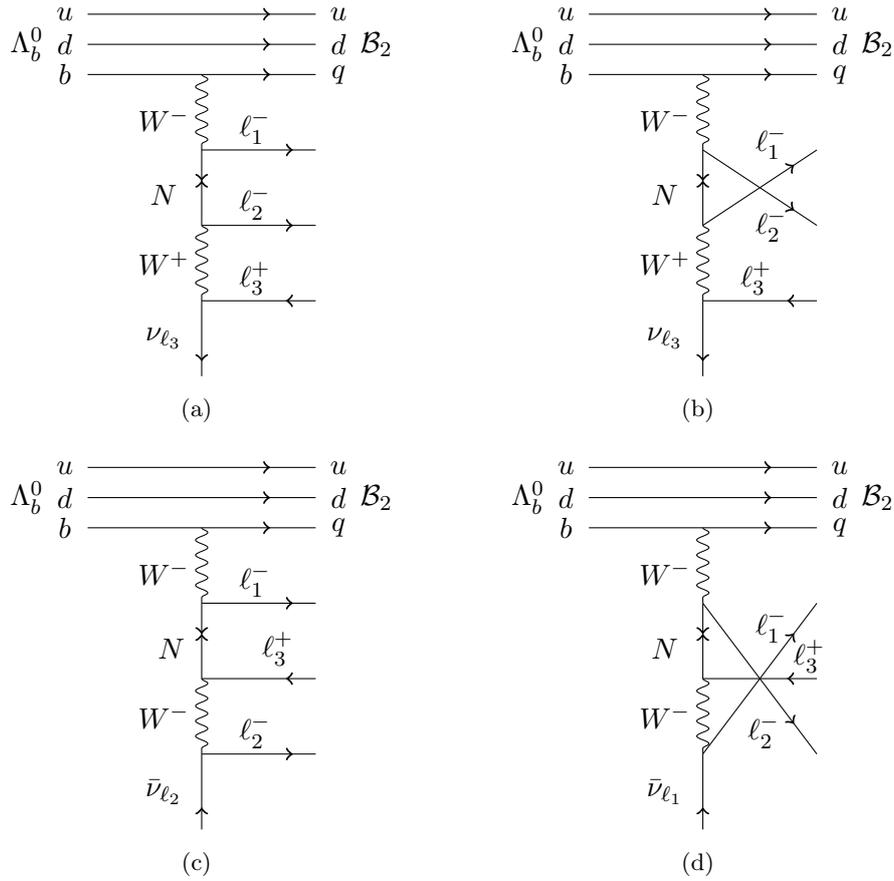
\begin{figure}[h!] 
\begin{center}
\subfigure[]{%
\begin{tikzpicture}
\draw[fermion,black] (0,0) --(3,0);
\draw[fermion,black] (0,0.4) --(3,0.4);
\draw[fermion,black] (0,0.8) --(3,0.8);
\draw[vector,black] (1.5,0) --(1.5,-1);
\draw[fermion,black] (1.5,-1) --(3,-1);
\draw[majoranafermion,black] (1.5,-1) --(1.5,-2);
\draw[fermion,black] (1.5,-2) --(3,-2);
\draw[vector,black] (1.5,-2) --(1.5,-3);
\draw[antifermion,black] (1.5,-3) --(3,-3);
\draw[fermion,black] (1.5,-3) --(1.5,-4);
\node at (-0.3,0) {$b$};
\node at (3.3,0) {$q$};
\node at (-0.3,0.4) {$d$};
\node at (3.3,0.4) {$d$};
\node at (-0.3,0.8) {$u$};
\node at (3.3,0.8) {$u$};
\node at (1, -.6) {$W^-$};
\node at (2.2, -0.7) {$\ell_1^-$};
\node at (1, -1.6) {$N$};
\node at (2.2, -1.7) {$\ell_2^-$};
\node at (1, -2.5) {$W^+$};
\node at (2.2, -2.7) {$\ell_3^+$};
\node at (1, -3.5) {$\nu_{\ell_3}$};
\node at (-0.8,0.4) {$\Lambda_b^0$};
\node at (3.8,0.4) {$\mathcal{B}_2$};
\end{tikzpicture}}
\hspace*{30pt}
\subfigure[]{%
\begin{tikzpicture}
\draw[fermion,black] (0,0) --(3,0);
\draw[fermion,black] (0,0.4) --(3,0.4);
\draw[fermion,black] (0,0.8) --(3,0.8);
\draw[vector,black] (1.5,0) --(1.5,-1);
\draw[fermion,black] (1.5,-1) --(3,-2);
\draw[majoranafermion,black] (1.5,-1) --(1.5,-2);
\draw[fermion,black] (1.5,-2) --(3,-1);
\draw[vector,black] (1.5,-2) --(1.5,-3);
\draw[antifermion,black] (1.5,-3) --(3,-3);
\draw[fermion,black] (1.5,-3) --(1.5,-4);
\node at (-0.3,0) {$b$};
\node at (3.3,0) {$q$};
\node at (-0.3,0.4) {$d$};
\node at (3.3,0.4) {$d$};
\node at (-0.3,0.8) {$u$};
\node at (3.3,0.8) {$u$};
\node at (1, -.6) {$W^-$};
\node at (2.4, -0.9) {$\ell_1^-$};
\node at (1, -1.6) {$N$};
\node at (2.4, -2) {$\ell_2^-$};
\node at (1, -2.5) {$W^{+}$};
\node at (2.2, -2.7) {$\ell_3^+$};
\node at (1, -3.5) {$\nu_{\ell_3}$};
\node at (-0.8,0.4) {$\Lambda_b^0$};
\node at (3.8,0.4) {$\mathcal{B}_2$};
\end{tikzpicture}}
\hspace*{30pt}
\subfigure[]{%
\begin{tikzpicture}
\draw[fermion,black] (0,0) --(3,0);
\draw[fermion,black] (0,0.4) --(3,0.4);
\draw[fermion,black] (0,0.8) --(3,0.8);
\draw[vector,black] (1.5,0) --(1.5,-1);
\draw[fermion,black] (1.5,-1) --(3,-1);
\draw[majoranafermion,black] (1.5,-1) --(1.5,-2);
\draw[antifermion,black] (1.5,-2) --(3,-2);
\draw[vector,black] (1.5,-2) --(1.5,-3);
\draw[fermion,black] (1.5,-3) --(3,-3);
\draw[antifermion,black] (1.5,-3) --(1.5,-4);
\node at (-0.3,0) {$b$};
\node at (3.3,0) {$q$};
\node at (-0.3,0.4) {$d$};
\node at (3.3,0.4) {$d$};
\node at (-0.3,0.8) {$u$};
\node at (3.3,0.8) {$u$};
\node at (1, -.6) {$W^-$};
\node at (2.2, -0.7) {$\ell_1^-$};
\node at (1.1, -1.6) {$N$};
\node at (2.5, -1.6) {$\ell_3^+$};
\node at (1, -2.5) {$W^{-}$};
\node at (2.2, -2.7) {$\ell_2^-$};
\node at (1, -3.5) {$\bar{\nu}_{\ell_2}$};
\node at (-0.8,0.4) {$\Lambda_b^0$};
\node at (3.8,0.4) {$\mathcal{B}_2$};
\end{tikzpicture}}
\hspace*{30pt}
\subfigure[]{%
\begin{tikzpicture}
\draw[fermion,black] (0,0) --(3,0);
\draw[fermion,black] (0,0.4) --(3,0.4);
\draw[fermion,black] (0,0.8) --(3,0.8);
\draw[vector,black] (1.5,0) --(1.5,-1);
\draw[fermion,black] (1.5,-1) --(3,-3);
\draw[majoranafermion,black] (1.5,-1) --(1.5,-2);
\draw[antifermion,black] (1.5,-2) --(3,-2);
\draw[vector,black] (1.5,-2) --(1.5,-3);
\draw[fermion,black] (1.5,-3) --(3,-1);
\draw[antifermion,black] (1.5,-3) --(1.5,-4);
\node at (-0.3,0) {$b$};
\node at (3.3,0) {$q$};
\node at (-0.3,0.4) {$d$};
\node at (3.3,0.4) {$d$};
\node at (-0.3,0.8) {$u$};
\node at (3.3,0.8) {$u$};
\node at (1, -.6) {$W^-$};
\node at (2.4, -1.3) {$\ell_1^-$};
\node at (1, -1.6) {$N$};
\node at (2.9, -1.7) {$\ell_3^+$};
\node at (1, -2.5) {$W^-$};
\node at (2.3, -2.7) {$\ell_2^-$};
\node at (1., -3.5) {$\bar{\nu}_{\ell_1}$};
\node at (-0.8,0.4) {$\Lambda_b^0$};
\node at (3.8,0.4) {$\mathcal{B}_2$};
\end{tikzpicture}}
\hspace*{30pt}
\caption{Figure (a) and (b) represent the LNV processes for direct and crossed channel diagram respectively. Figure (c) and (d) represent the LNC processes for direct and crossed channel Feynman diagram respectively. Here $q$ is the charm and up quark for $\mathcal{B}_2 = \Lambda_c^+ $ and proton ($p^+$) respectively.\label{fig:feynman}}
\end{center}
\end{figure}

Our interest is to calculate  $\Lambda_b^0\to (\Lambda_c^+,p^+)\ell_1^-\ell_2^-\ell_3^+\nu$ decays and their conjugate modes. Depending on the flavor of the neutrinos that are unobserved, the decays can be either LNV or LNC. For example, $\Lambda_b^0 \to (\Lambda_c^+,p^+) \ell_1^- \ell_2^-\ell_3^+\nu$ is LNV if $\nu$ is $\nu_{\ell_3}$, but it is LNC if $\nu$ is $\bar{\nu}_{\ell_2}$.  The LNV decays occur due to Majorana neutrinos whereas LNC decays can happen either due to Majorana or a Dirac. At the LHC, the reconstruction efficiency of tau leptons are is smaller than muon and electrons. Moreover, the tau leptons in the final state are also expected to give severe phase space suppressions. For these reasons, we neglect tau leptons. To reduce the dominant electromagnetic backgrounds we also avoid the events where two leptons of opposite charges are present. Therefore, we will study only the following combinations of leptons in the $\Lambda_b^0 \rightarrow (\Lambda_c^+,p^+) \ell_1^- \ell_2^- \ell_3^+ \nu$ decays. Depending on the flavor of the unobserved neutrinos the final states can be either LNV or LNC.
\begin{eqnarray}
& {\rm LNV}:\,\,\ (i)\,\ \ell_1=\ell_2=\mu,\,\,\ \ell_3=e,\,\ \nu=\nu_{\ell_3};\,\ (ii)\,\ \ell_1=\ell_2=e,\,\,\ \ell_3=\mu,\,\ \nu=\nu_{\ell_3}\nonumber\ \\
	& {\rm LNC}:\,\,\ (i)\,\ \ell_1=\ell_2=\mu,\,\,\ \ell_3=e,\,\ \nu=\bar{\nu}_{\ell_2};\,\ (ii)\,\ \ell_1=\ell_2=e,\,\,\ \ell_3=\mu,\,\ \nu=\bar{\nu}_{\ell_2}\nn
\end{eqnarray}
To clarify our convention, when the decay is LNV, the $\ell_1,\ell_2$ are assumed attached to the vertexes of the sterile neutrino and the $\ell_3\nu$ pair comes from a $W$. When the decay is LNC, the $\ell_1,\ell_3$ are assumed attached to the vertices of the sterile neutrino and the $\ell_2\nu$ pair comes from a $W$. The Feynman diagrams for these relevant processes are shown in figure $\ref{fig:feynman}$. The dominant contributions come from two "s-channel" topologies, direct(D), and cross-channel. There are no ``t-channel'' diagrams for these processes.  The decay rates are appreciable if the sterile neutrinos have kinematically allowed masses 
\begin{equation}
m_{\ell_2} + m_{\ell_3} < m_{N_j} < (\mB{1}-\mB{2}-m_{\ell_1})\, ,\quad \text{and/or}\quad m_{\ell_1} + m_{\ell_3} < m_{N_j} < (\mB{1}-\mB{2}-m_{\ell_2})
\end{equation}
%
%

\subsection{$\mathcal{B}_1\to\mathcal{B}_2\ell_1N\to \mathcal{B}_2\ell_1\ell_2\ell_3\nu$ rates }
The five-body decay can be viewed as a two-step process, production of the heavy neutrino in the semi-leptonic decay $\mathcal{B}_1\to\mathcal{B}_2\ell_1N$, followed by the decay of the heavy neutrino $N\to \ell_2\ell_3\nu$. With a Majorana neutrino being exchanged, the process will be LNV and the decay rates can be written as \cite{Das:2021prm}  
\begin{align}
\label{eq:b1rate}
\Gamma_{\mathcal{B}_1}^{\rm LNV} &\equiv \Gamma (\mathcal{B}_1\rightarrow \mathcal{B}_2 \ell_1^-  \ell_2^- \ell_3^+ \nu_{\ell_3}) =  (2-\delta_{\ell_1 \ell_2})\frac{1}{2!}\Bigg[|V_{\ell_1 N_1}|^2 |V_{\ell_2 N_1}|^2 \bigg( \widehat{\Gamma}(DD^\ast)_{11} + \widehat{\Gamma}(CC^\ast)_{11}\bigg) \, \nn\\ &+ |V_{\ell_1 N_2}|^2 |V_{\ell_2 N_2}|^2 \bigg( \widehat{\Gamma}(DD^\ast)_{22} +  \widehat{\Gamma}(CC^\ast)_{22}\bigg) \, \nn\\ & + 2 |V_{\ell_1 N_1}| |V_{\ell_2 N_1}| |V_{\ell_1 N_2}| |V_{\ell_2 N_2}| \bigg\{ \cos(\theta_{21}^{\rm LNV})\Big(\re\widehat{\Gamma}(DD^\ast)_{12} + \re\widehat{\Gamma}(CC^\ast)_{12} \Big) \,\nn\\
&+ \sin(\theta_{21}^{\rm LNV}) \Big(\im\widehat{\Gamma}(DD^\ast)_{12} + \im\widehat{\Gamma}(CC^\ast)_{12} \Big)  \bigg\} 
\Bigg]\,,\\
\label{eq:b1barrate}
\Gamma_{\mathcal{\overline{B}}_1}^{\rm LNV} &\equiv \Gamma (\mathcal{\bar{B}}_1\rightarrow \mathcal{\bar{B}}_2 \ell_1^+  \ell_2^+ \ell_3^- \bar{\nu}_{\ell_3}) =  (2-\delta_{\ell_1 \ell_2})\frac{1}{2!}\Bigg[|V_{\ell_1 N_1}|^2 |V_{\ell_2 N_1}|^2 \bigg( \widehat{\Gamma}(DD^\ast)_{11} + \widehat{\Gamma}(CC^\ast)_{11}\bigg) \, \nn\\ &+ |V_{\ell_1 N_2}|^2 |V_{\ell_2 N_2}|^2 \bigg( \widehat{\Gamma}(DD^\ast)_{22} +  \widehat{\Gamma}(CC^\ast)_{22}\bigg) \, \nn\\ & + 2 |V_{\ell_1 N_1}| |V_{\ell_2 N_1}| |V_{\ell_1 N_2}| |V_{\ell_2 N_2}| \bigg\{ \cos(\theta_{21}^{\rm LNV})\Big(\re\widehat{\Gamma}(DD^\ast)_{12} + \re\widehat{\Gamma}(CC^\ast)_{12} \Big) \,\nn\\
& - \sin(\theta_{21}^{\rm LNV}) \Big(\im\widehat{\Gamma}(DD^\ast)_{12} + \im\widehat{\Gamma}(CC^\ast)_{12} \Big)  \bigg\} 
\Bigg].
\end{align}
Based on the convention \eqref{eq:convention} 
\begin{align}
\begin{split}
\theta_{21}^{\rm LNV}&= \arg(V_{\ell_1N_2})+\arg(V_{\ell_2N_2})-\arg(V_{\ell_1N_1})-\arg(V_{\ell_2N_1})\, ,\\
&=(\phi_{\ell_1 N_2}+\phi_{\ell_2 N_2}-\phi_{\ell_1 N_1}-\phi_{\ell_2 N_1}),\, .
\end{split}
\end{align}

The LNC finals state can result either due to Majorana or Dirac neutrinos. The expression of LNC rates are 
\begin{align}
\label{eq:b3rate}
\Gamma_{\mathcal{B}_1}^{\rm LNC} &\equiv \Gamma (\mathcal{B}_1\rightarrow \mathcal{B}_2 \ell_1^-  \ell_2^- \ell_3^+ \bar{\nu}_{\ell_2}) =  (2-\delta_{\ell_1 \ell_2})\frac{1}{2!}\Bigg[|V_{\ell_1 N_1}|^2 |V_{\ell_3 N_1}|^2 \bigg( \widehat{\Gamma}(DD^\ast)_{11} + \kappa_1^2 \widehat{\Gamma}(CC^\ast)_{11}\bigg) \, \nn\\ &+ |V_{\ell_1 N_2}|^2 |V_{\ell_3 N_2}|^2 \bigg( \widehat{\Gamma}(DD^\ast)_{22} +  \kappa_2^2 \widehat{\Gamma}(CC^\ast)_{22}\bigg) \, \nn\\ & + 2 |V_{\ell_1 N_1}| |V_{\ell_3 N_1}| |V_{\ell_1 N_2}| |V_{\ell_3 N_2}| \bigg\{ \cos(\theta_{21}^{\rm LNC})\Big(\re\widehat{\Gamma}(DD^\ast)_{12} + \kappa_1\kappa_2\re\widehat{\Gamma}(CC^\ast)_{12} \Big) \,\nn\\
&+ \sin(\theta_{21}^{\rm LNC}) \Big(\im\widehat{\Gamma}(DD^\ast)_{12} + \kappa_1\kappa_2\im\widehat{\Gamma}(CC^\ast)_{12} \Big)  \bigg\} 
\Bigg]\,,\\
\label{eq:b4barrate}
\Gamma_{\mathcal{\overline{B}}_1}^{\rm LNC} &\equiv \Gamma (\mathcal{\bar{B}}_1\rightarrow \mathcal{\bar{B}}_2 \ell_1^+  \ell_2^+ \ell_3^- \nu_{\ell_2}) =  (2-\delta_{\ell_1 \ell_2})\frac{1}{2!}\Bigg[|V_{\ell_1 N_1}|^2 |V_{\ell_3 N_1}|^2 \bigg( \widehat{\Gamma}(DD^\ast)_{11} + \kappa_1^2 \widehat{\Gamma}(CC^\ast)_{11}\bigg) \, \nn\\ &+ |V_{\ell_1 N_2}|^2 |V_{\ell_3 N_2}|^2 \bigg( \widehat{\Gamma}(DD^\ast)_{22} +  \kappa_2^2 \widehat{\Gamma}(CC^\ast)_{22}\bigg) \, \nn\\ & + 2 |V_{\ell_1 N_1}| |V_{\ell_3 N_1}| |V_{\ell_1 N_2}| |V_{\ell_3 N_2}| \bigg\{ \cos(\theta_{21}^{\rm LNC})\Big(\re\widehat{\Gamma}(DD^\ast)_{12} + \kappa_1\kappa_2\re\widehat{\Gamma}(CC^\ast)_{12} \Big) \,\nn\\
& - \sin(\theta_{21}^{\rm LNC}) \Big(\im\widehat{\Gamma}(DD^\ast)_{12} +\kappa_1\kappa_2 \im\widehat{\Gamma}(CC^\ast)_{12} \Big)  \bigg\} 
\Bigg]\, .
\end{align}
The expressions of $\widehat{\Gamma}(XX^\ast)$ are different for LNV and LNC rates and have been derived in Appendix \ref{sec:fivebody}. The factor 1/2! in the decay rates is the combinatorial factor when to account for the scenario when two charged leptons in the final state are the same. The CP-odd phase $\theta^{\rm LNC}_{21}$, and $\kappa_{1,2}$ are defined as
\begin{align}
\begin{split}
\theta_{21}^{\rm LNC}&= \arg(V_{\ell_1N_2})-\arg(V_{\ell_3N_2})-\arg(V_{\ell_1N_1})+\arg(V_{\ell_3N_1})\, ,\\
&=(\phi_{\ell_1 N_2}-\phi_{\ell_3 N_2}-\phi_{\ell_1 N_1}+\phi_{\ell_3 N_1}),\,\\
\kappa_1 &= \frac{|V_{\ell_2 N_1 }|}{|V_{\ell_1 N_1}| },\,\,\,\ \kappa_2 = \frac{|V_{\ell_2 N_2 }|}{|V_{\ell_1 N_2}| } 
\end{split}
\end{align}
In general, the CP-odd phases $\theta_{21}^{\rm LNV}$ and $\theta_{21}^{\rm LNC}$ are different and $\theta_{21}^{\rm LNV}$ is not equal to $\theta_{21}^{\rm LNC}$ by the replacement $\ell_2\to \ell_3$.

\begin{figure}[h!]
	\begin{center}
		\includegraphics[scale=0.15]{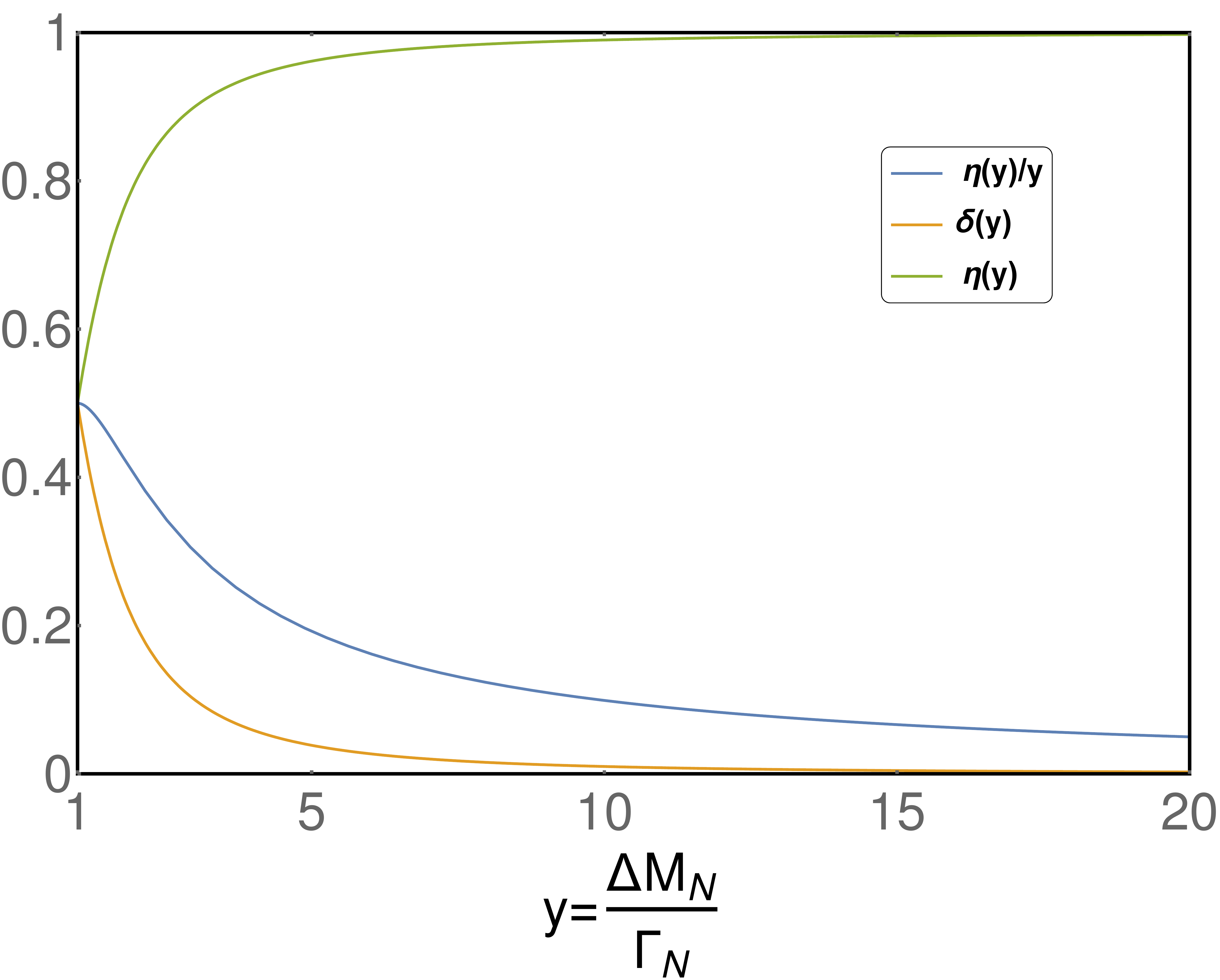}
		\caption{Variation of overlapping function $\eta(y)$, $\eta(y)/y$ and $\delta(y)$ with $y= \frac{\Delta M_{N}}{\Gamma_N }$. \label{fig:eta}}
	\end{center}
\end{figure}

In the decay rate formulas given above, contributions coming from $D-C$ channel interference have been neglected as they are negligibly small. To give some physical interpretation, up to the heavy-to-light mixing elements, the diagonal terms $\widehat{\Gamma}(XX^\ast)_{ij}, X=D,C$ are the decay rates involving $i^{\rm th}$ neutrino in the $X$ channel and $j^{\rm th}$ neutrino in the conjugate channel. The on-shell assumption $\Gamma_{N_j}\ll m_{N_j}$ is always valid in our case, and the diagonal elements can be calculated analytically as
%
\begin{eqnarray}\label{eq:aux1}
\widehat{\Gamma}(DD^\ast)_{jj} =  \overline{\Gamma}(\mathcal{B}_1\to \mathcal{B}_2 \ell_1 N_j) \times \frac{\overline{\Gamma}(N_j\to \ell_2\ell_3\nu)}{\Gamma_{N_j}},\,\ \widehat{\Gamma}(CC^\ast)_{jj}=\widehat{\Gamma}(DD^\ast)_{jj}(\ell_1 \leftrightarrow \ell_2).\nn\ \\
\end{eqnarray}
where we have adopted the notation $\Gamma(\mathcal{B}_1\to \mathcal{B}_2 \ell_1 N_j) = |V_{\ell_1 N}|^2 \overline{\Gamma}(\mathcal{B}_1\to \mathcal{B}_2 \ell_1 N_j)$ and $\Gamma(N_j\to \ell_2\ell_3\nu) = |V_{(\ell_2/\ell_3) N}|^2 \overline{\Gamma}(N_j\to \ell_2\ell_3\nu)$. As the neutrinos are almost degenerate we can assume that 
\begin{eqnarray}
\begin{split}\label{eq:aux2}
\overline{\Gamma}(\mathcal{B}_1\to\mathcal{B}_2 \ell_1 N_1)&= \overline{\Gamma}(\mathcal{B}_1\to\mathcal{B}_2 \ell_1 N_2)\equiv \overline{\Gamma}(\mathcal{B}_1\to\mathcal{B}_2 \ell_1 N)\,,\\
\overline{\Gamma}(N_1 \to \ell_2\ell_3\nu) &= \overline{\Gamma}(N_2 \to \ell_2\ell_3\nu)\equiv \overline{\Gamma}(N \to \ell_2\ell_3\nu)\,.
\end{split}
\end{eqnarray}
The differential expressions of the normalized three body decay width $d\overline{\Gamma}(\mathcal{B}_1\to\mathcal{B}_2 \ell_1 N)$ are given in Appendix \ref{sec:matrix} and related three body decay kinematics for this relevant process are given in Appendix \ref{sec:kinematics}. The explicit expression of normalized decay width of heavy sterile neutrino $N$ without heavy to light mixing element, $\overline{\Gamma}(N \to \ell_2\ell_3\nu)$ is taken from \cite{Cvetic:2017vwl}. The off-diagonal elements $\re\widehat{\Gamma}(XX^\ast)_{12}$ and $\im\widehat{\Gamma}(XX^\ast)_{12}$ are the ones where the two neutrinos interfere making them sensitive to the mass difference $\Delta M_N$, and hence, on the assumption $\Gamma_{N_j}\ll \Delta M_N$. This sensitivity can be parametrized in terms of $\delta(y)$ and $\eta(y)$ \cite{Das:2021prm}
\begin{eqnarray}
\begin{split}\label{eq:aux3}
\frac{\re \widehat{\Gamma}(XX^\ast)_{12}}{\widehat{\Gamma}(XX^\ast)_{jj}} &= 2 \delta(y)\frac{\Gamma_{N_j}}{\Gamma_{N_1}+\Gamma_{N_2}}\,  ,\\
\frac{\im \widehat{\Gamma}(XX^\ast)_{12}}{\widehat{\Gamma}(XX^\ast)_{jj}} &= 2 \frac{\eta(y)}{y}\frac{\Gamma_{N_j}}{\Gamma_{N_1}+\Gamma_{N_2}},\,\,\ X=C,D; j=1,2
\end{split}
\end{eqnarray}
where $y = \Delta M_N/\Gamma_N$, and $\Gamma_N = (\Gamma_{N_1} + \Gamma_{N_2})/2$ is the average decay width. Numerically we find 
\begin{equation}\label{eq:aux4}
\eta(y) = \frac{y^2}{y^2+1}\, ,\quad \delta(y) = \frac{1}{y^2+1}\, .
\end{equation}
which is in agreement with similar decays in mesons \cite{Cvetic:2020lyh}. Physically, $\delta(y)$ and $\eta(y)$ are correction terms for non-negligible overlap of the two neutrinos. In figure $\ref{fig:eta}$ we show $\eta$ and $\delta$ as a function of $y$. $\re \widehat{\Gamma}(XX^\ast)_{12}$ and $\im \widehat{\Gamma}(XX^\ast)_{12}$ are also sources of a CP-even phase $\Delta\xi = \xi_1-\xi_2$ where 
\begin{equation}
\tan\xi_1 = \frac{m_{N_1} \Gamma_{N_1}}{k_N^2 - m_{N_1}^2},\,\,\ \tan\xi_2 = \frac{m_{N_2} \Gamma_{N_2}}{k_N^2 - m_{N_2}^2}\, .\quad 
\end{equation}
where $k_N$ is the momentum of the intermediate neutrinos. The cosine and sine of the angles $\Delta\xi$ come from the $\re \widehat{\Gamma}(XX^\ast)_{12}$ and $\im \widehat{\Gamma}(XX^\ast)_{12}$ parts, respectively \cite{Das:2021prm}. 

We will use the equations \eqref{eq:aux1}-\eqref{eq:aux3} in the subsequent sections to simplify the decay rates.

\subsection{Neutrino oscillation}
In the previous expressions of decay widths, two effects are missing, acceptance factor $P_N$, which is a probability that the sterile neutrino decays within the detector in its flight, and the oscillation between $N_1$ and $N_2$ states. The effect of heavy neutrino oscillation is that the decay rates depend on the length of flight $L$ of the heavy neutrinos between production and decay vertex \cite{Cvetic:2015ura}. The acceptance factors is given by
\begin{eqnarray}\label{eq:probability}
P_{N_j}&=&1-\exp\Big(-\frac{t}{\tau_{N_j}\gamma_{N_j}}\Big)=1-\exp\Big(-\frac{L \Gamma_{N_j}}{\beta_{N_j}\gamma_{N_j}}\Big)\,,
\end{eqnarray}
where $\tau_{N_j}$ is the lifetime of $N_j$, $\beta_{N_j}$ is the velocity in the lab frame, and $\gamma_{N_j} = (1-\beta_{N_j}^2)^{-1/2}$. Though in practice $P_{N_j}$ can be different for two neutrinos, they being degenerate in mass we assume $\Gamma_{N_1}\approx\Gamma_{N_2}=\Gamma_N$, $\beta_{N_1} = \beta_{N_2} \equiv \beta_{N}$. The effective decay widths including the acceptance factor are
\begin{eqnarray}\label{eq:GammaEff}
\Gamma_{\mathcal{B}_1/\bar{\mathcal{B}}_1}^{\rm LNV(LNC),eff} 
&=&\bigg( 1-\exp\Big(-\frac{L \Gamma_{N}^{\rm M/D}}{\beta_{N}\gamma_{N}}\Big)  \bigg) \Gamma_{\mathcal{B}_1/\bar{\mathcal{B}}_1}^{\rm LNV(LNC)}
\end{eqnarray}
where the expressions of $\Gamma_{\mathcal{B}_1/\bar{\mathcal{B}}_1}^{\rm LNV(LNC)}$ are given in the previous section, and $\Gamma_N^{\rm M/D}$ indicate decay rates of Majorana (M) or Dirac (D) neutrinos. For the rest of this section up to the end, we will show the derivations only for the LNV mode. For the LNC mode, only the final expression will be given but the steps of derivations are similar. The effective differential decay width with respect to $L$ for LNV mode is given as 
\begin{eqnarray}\label{eq:dGammaEff1}
\frac{d \Gamma_{\mathcal{B}_1/\bar{\mathcal{B}}_1}^{\rm LNV,eff}}{dL} &=& \frac{ \Gamma_{N}^{\rm M}}{\beta_{N}\gamma_{N}} \exp\bigg(-\frac{L \Gamma_{N}^{\rm M}}{\beta_{N}\gamma_{N}}\  \bigg) \bar{\Gamma} (\mathcal{B}_1\rightarrow \mathcal{B}_2 \ell_1^- N) \bar{\Gamma} (N\rightarrow \ell_2^- \ell_3^+ \nu_{\ell_3}) \times \bigg\{\frac{|V_{\ell_1 N_1}|^2 |V_{\ell_2 N_1}|^2}{\Gamma_{N_1}^{\rm M}} \nonumber \\
&+& \frac{|V_{\ell_1 N_2}|^2 |V_{\ell_2 N_2}|^2}{\Gamma_{N_2}^{\rm M}} + \frac{4\big(|V_{\ell_1 N_1}||V_{\ell_2 N_1}||V_{\ell_1 N_2}||V_{\ell_2 N_2}|\big)}{\Gamma_{N_1}^{\rm M}+\Gamma_{N_2}^{\rm M}}\Big(\delta(y)\cos(\theta_{21}^{\rm LNV}) \nonumber \\ &\pm & \frac{\eta(y)}{y}\sin(\theta_{21}^{\rm LNV} )\Big) \bigg\}\,  .~~
\end{eqnarray}
To obtain \eqref{eq:dGammaEff1} from \eqref{eq:GammaEff} we have used \eqref{eq:aux1}, \eqref{eq:aux2} and \eqref{eq:aux3}. 

We now consider a scenario where the neutrinos are degenerate, ie $\Delta M_N\ll m_{N_{i,j}}$, but $\Delta M_N \gg \Gamma_N$. In this case $y\gg 1$ and hence the functions $\delta(y)$ and $\eta(y)/y$ are negligibly small. Physically, this corresponds to a scenario where the overlap between the two neutrinos is negligibly small. In the $y\gg 1$ limit the effective differential decay rates for LNV modes with respect to $L$ are
\begin{eqnarray}\label{eq:dGammaEff3}
\frac{d \Gamma_{\mathcal{B}_1/\bar{\mathcal{B}}_1}^{\rm LNV,eff}}{dL} &=& \frac{ \Gamma_{N}^{\rm M}}{\beta_{N}\gamma_{N}} \exp\bigg(-\frac{L \Gamma_{N}^{\rm M}}{\beta_{N}\gamma_{N}}\  \bigg) \bar{\Gamma} (\mathcal{B}_1\rightarrow \mathcal{B}_2 \ell_1^- N) \bar{\Gamma} (N\rightarrow \ell_2^- \ell_3^+ \nu_{\ell_3}) \times \bigg\{\frac{|V_{\ell_1 N_1}|^2 |V_{\ell_2 N_1}|^2}{\Gamma_{N_1}^{\rm M}} \nonumber \\
&+& \frac{|V_{\ell_1 N_2}|^2 |V_{\ell_2 N_2}|^2}{\Gamma_{N_2}^{\rm M}} \bigg\}\, .~~
\end{eqnarray}
To introduce effects of $N_1-N_2$ oscillation we note that while a neutrino travels from the production vertex to the decay vertex, it picks up a phase $\exp(-ip_{N_j}\cdot z)$ where $p_{N_j}$ is the four momentum of $N_j$, $z = (t,0,0,L)$ is the space-time separation between the two vertices, and $L\approx \beta_N t$ is the length traveled by the neutrinos between production and decay vertex \cite{Cohen:2008qb}. This amplitudes is 
\begin{equation}
\mathcal{A}^{\rm LNV}_{\rm osc}(\mathcal{B}_1 \rightarrow \mathcal{B}_2 \ell_1^- \ell_2^-\ell_3^+\nu_{\ell_3};L) \sim V_{\ell_1 N_1} V_{\ell_2 N_1} \exp(-ip_{N_1}\cdot z) + V_{\ell_1 N_2} V_{\ell_2 N_2} \exp(-ip_{N_2}\cdot z)\, .
\end{equation}
The expressions for $\mathcal{A}^{\rm LNV}_{\rm osc}(\mathcal{\bar{B}}_1 \rightarrow \mathcal{\bar{B}}_2 \ell_1^+ \ell_2^+\ell_3^-\bar{\nu}_{\ell_3};L)$ is obtained by taking complex conjugate of the $V_{\ell N}$ elements in the above formulas.  The differential with respect to $L$ of the decay rates that include oscillation effect is proportional to the modulus squared of the amplitudes and we can write
\begin{align}\label{eq:ampLNV}
&\frac{d }{dL} \Gamma^{\rm LNV}_{\rm osc}(\mathcal{B}_1 \rightarrow \mathcal{B}_2 \ell_1^- \ell_2^-\ell_3^+\nu_{\ell_3};L) \sim \big|\mathcal{A}^{\rm LNV}_{\rm osc}(\mathcal{B}_1 \rightarrow \mathcal{B}_2 \ell_1^- \ell_2^-\ell_3^+\nu_{\ell_3};L)\big|^2 \nn\ \\ 
&\sim \sum_{i=1}^2 |V_{\ell_1 N_i}|^2|V_{\ell_2 N_i}|^2 + 2 \re \Big(V_{\ell_1 N_1}V_{\ell_2 N_1}V^\ast_{\ell_1 N_2} V^\ast_{\ell_2 N_2} \exp[i(p_{N_2}-p_{N_1}).z] \Big)
\end{align}
The superscript ``osc'' indicates that the oscillation effect is included in the effective differential decay width. Comparing with the the differential decay widths \eqref{eq:dGammaEff3} and \eqref{eq:ampLNV} we get the differential expressions of decay rates that includes oscillation as
\begin{eqnarray}\label{eq:effOscLNV}
\frac{d \Gamma_{\mathcal{B}_1/\bar{\mathcal{B}}_1}^{\rm LNV,osc}}{dL} &=& \frac{ \Gamma_{N}^{\rm M}}{\beta_{N}\gamma_{N}} \exp\bigg(-\frac{L \Gamma_{N}^{\rm M}}{\beta_{N}\gamma_{N}}\  \bigg) \bar{\Gamma} (\mathcal{B}_1\rightarrow \mathcal{B}_2 \ell_1^- N) \bar{\Gamma} (N\rightarrow \ell_2^- \ell_3^+ \nu_{\ell_3}) \times \bigg\{\frac{|V_{\ell_1 N_1}|^2 |V_{\ell_2 N_1}|^2}{\Gamma_{N_1}^{\rm M}} \nonumber \\
&+& \frac{|V_{\ell_1 N_2}|^2 |V_{\ell_2 N_2}|^2}{\Gamma_{N_2}^{\rm M}} +  \frac{4  |V_{\ell_1 N_1}||V_{\ell_2 N_1}||V_{\ell_1 N_2}|| V_{\ell_2 N_2}|}{\Gamma_{N_1}^{\rm M}+\Gamma_{N_2}^{\rm M}} \cos\Big(\frac{2\pi L}{L_{\rm osc}}   \mp \theta_{21}^{\rm LNV}\Big) \bigg\}\, .
\end{eqnarray} 
To simplify the above equations we have used $(p_{N_2}-p_{N_1}).z=2\pi L/L_{\rm osc}$, and $L_{\rm osc}=2\pi\beta_N\gamma_N/\Delta M_N= 2\pi\beta_N\gamma_N/(y \Gamma_N)$. Combining \eqref{eq:dGammaEff1} and \eqref{eq:effOscLNV} we get the complete expression of differential decay width that includes both the acceptance factor and the oscillation effects 
\begin{eqnarray}
\frac{d \Gamma_{\mathcal{B}_1/\bar{\mathcal{B}}_1}^{\rm LNV,osc}}{dL} &=& \frac{ \Gamma_{N}^{\rm M}}{\beta_{N}\gamma_{N}} \exp\bigg(-\frac{L \Gamma_{N}^{\rm M}}{\beta_{N}\gamma_{N}}\  \bigg) \bar{\Gamma} (\mathcal{B}_1\rightarrow \mathcal{B}_2 \ell_1^- N) \bar{\Gamma} (N\rightarrow \ell_2^- \ell_3^+ \nu_{\ell_3}) \times \bigg\{\frac{|V_{\ell_1 N_1}|^2 |V_{\ell_2 N_1}|^2}{\Gamma_{N_1}^{\rm M}} \nonumber \\
&+& \frac{|V_{\ell_1 N_2}|^2 |V_{\ell_2 N_2}|^2}{\Gamma_{N_2}^{\rm M}} + \frac{4\big(|V_{\ell_1 N_1}||V_{\ell_2 N_1}||V_{\ell_1 N_2}||V_{\ell_2 N_2}|\big)}{\Gamma_{N_1}^{\rm M}+\Gamma_{N_2}^{\rm M}}\Big(\delta(y)\cos(\theta_{21}^{\rm LNV}) \pm \frac{\eta(y)}{y}\sin(\theta_{21}^{\rm LNV} )\Big)\nn\\\
&+& \frac{4}{\Gamma_{N_1}^{\rm M}+\Gamma_{N_2}^{\rm M}} |V_{\ell_1 N_1}||V_{\ell_2 N_1}||V_{\ell_1 N_2}|| V_{\ell_2 N_2}| \cos\Big(\frac{2\pi L}{L_{\rm osc}} \mp \theta_{21}^{\rm LNV}\Big) \bigg\}\, ,
\end{eqnarray} 
where we have used the equations \eqref{eq:aux1}-\eqref{eq:aux3}. 
Integrating the above equations with respect to $L$ we obtain 
\begin{eqnarray}\label{eq:finalLNV}
\Gamma_{\mathcal{B}_1/\bar{\mathcal{B}}_1}^{\rm LNV,osc} &\equiv & \Gamma (\mathcal{B}_1\rightarrow \mathcal{B}_2 \ell_1^-  \ell_2^- \ell_3^+ \nu_{\ell_3})
 =\bar{\Gamma} (\mathcal{B}_1\rightarrow \mathcal{B}_2 \ell_1^- N) \bar{\Gamma} (N\rightarrow \ell_2^- \ell_3^+ \nu_{\ell_3})\nonumber\\ &\times &\bigg\{ \bigg[\frac{|V_{\ell_1 N_1}|^2 |V_{\ell_2 N_1}|^2}{\Gamma_{N_1}^{\rm M}} + \frac{|V_{\ell_1 N_2}|^2 |V_{\ell_2 N_2}|^2}{\Gamma_{N_2}^{\rm M}} \nonumber \\
&+&\frac{4}{\Gamma_{N_1}^{\rm M}+\Gamma_{N_2}^{\rm M}}|V_{\ell_1 N_1}||V_{\ell_2 N_1}||V_{\ell_1 N_2}||V_{\ell_2 N_2}|\Big(\delta(y)\cos(\theta_{21}^{\rm LNV})\nonumber \\
& \pm & \frac{\eta(y)}{y}\sin(\theta_{21}^{\rm LNV} )\Big)\bigg] \Big( 1-e^{\frac{-L\Gamma_N^{\rm M}}{\beta_N\gamma_N}}\Big) \nonumber\\\
&+& \frac{4}{\Gamma_{N_1}^{\rm M}+\Gamma_{N_2}^{\rm M}}\frac{1}{1+y^2}|V_{\ell_1 N_1}||V_{\ell_2 N_1}||V_{\ell_1 N_2}||V_{\ell_2 N_2}|\bigg(e^{\frac{-L\Gamma_N^{\rm M}}{\beta_N\gamma_N}}\Big[y \sin\Big(\frac{2\pi L}{L_{osc}}\mp \theta_{21}^{\rm LNV}\Big)\nonumber\\\
&-&\cos\Big(\frac{2\pi L}{L_{osc}}\mp \theta_{21}^{\rm LNV}\Big)\Big]+\Big[\cos\big(\theta_{21}^{\rm LNV}\big)\pm y \sin\big(\theta_{21}^{\rm LNV}\big) \Big]\bigg)\bigg\}
\end{eqnarray}
Following similar steps, we give the final expression for LNC decay
\begin{eqnarray}\label{eq:finalLNC}
\Gamma_{\mathcal{B}_1/\bar{\mathcal{B}_1}}^{\rm F,LNC,osc} &\equiv & \Gamma (\mathcal{B}_1\rightarrow \mathcal{B}_2 \ell_1^-  \ell_2^- \ell_3^+ \bar{\nu}_{\ell_2})
=\bar{\Gamma} (\mathcal{B}_1\rightarrow \mathcal{B}_2 \ell_1^- N) \bar{\Gamma} (N\rightarrow \ell_2^- \ell_3^+ \bar{\nu}_{\ell_2})\nonumber\\ &\times &\bigg\{ \bigg[\frac{|V_{\ell_1 N_1}|^2 |V_{\ell_3 N_1}|^2}{\Gamma_{N_1}^{\rm F}} + \frac{|V_{\ell_1 N_2}|^2 |V_{\ell_3 N_2}|^2}{\Gamma_{N_2}^{\rm F}} \nonumber \\
&+&\frac{4}{\Gamma_{N_1}^{\rm F}+\Gamma_{N_2}^{\rm F}}|V_{\ell_1 N_1}||V_{\ell_3 N_1}||V_{\ell_1 N_2}||V_{\ell_3 N_2}|\Big(\delta(y)\cos(\theta_{21}^{\rm LNC})\nonumber \\
& \pm & \frac{\eta(y)}{y}\sin(\theta_{21}^{\rm LNC} )\Big)\bigg] \Big( 1-e^{\frac{-L\Gamma_N^{\rm F}}{\beta_N\gamma_N}}\Big) \nonumber\\\
&+& \frac{4}{\Gamma_{N_1}^{\rm F}+\Gamma_{N_2}^{\rm F}}\frac{1}{1+y^2}|V_{\ell_1 N_1}||V_{\ell_3 N_1}||V_{\ell_1 N_2}||V_{\ell_3 N_2}|\bigg(e^{\frac{-L\Gamma_N^{\rm F}}{\beta_N\gamma_N}}\Big[y \sin\Big(\frac{2\pi L}{L_{osc}}\mp \theta_{21}^{\rm LNC}\Big)\nonumber\\\
&-&\cos\Big(\frac{2\pi L}{L_{osc}}\mp \theta_{21}^{\rm LNC}\Big)\Big]+\Big[\cos\big(\theta_{21}^{\rm LNC}\big)\pm y \sin\big(\theta_{21}^{\rm LNC}\big) \Big]\bigg)\bigg\}\, .
\end{eqnarray} 
where F = M (Majorana) or D (Dirac).

\begin{figure}[h!]
	\begin{center}
		\includegraphics[scale=0.15]{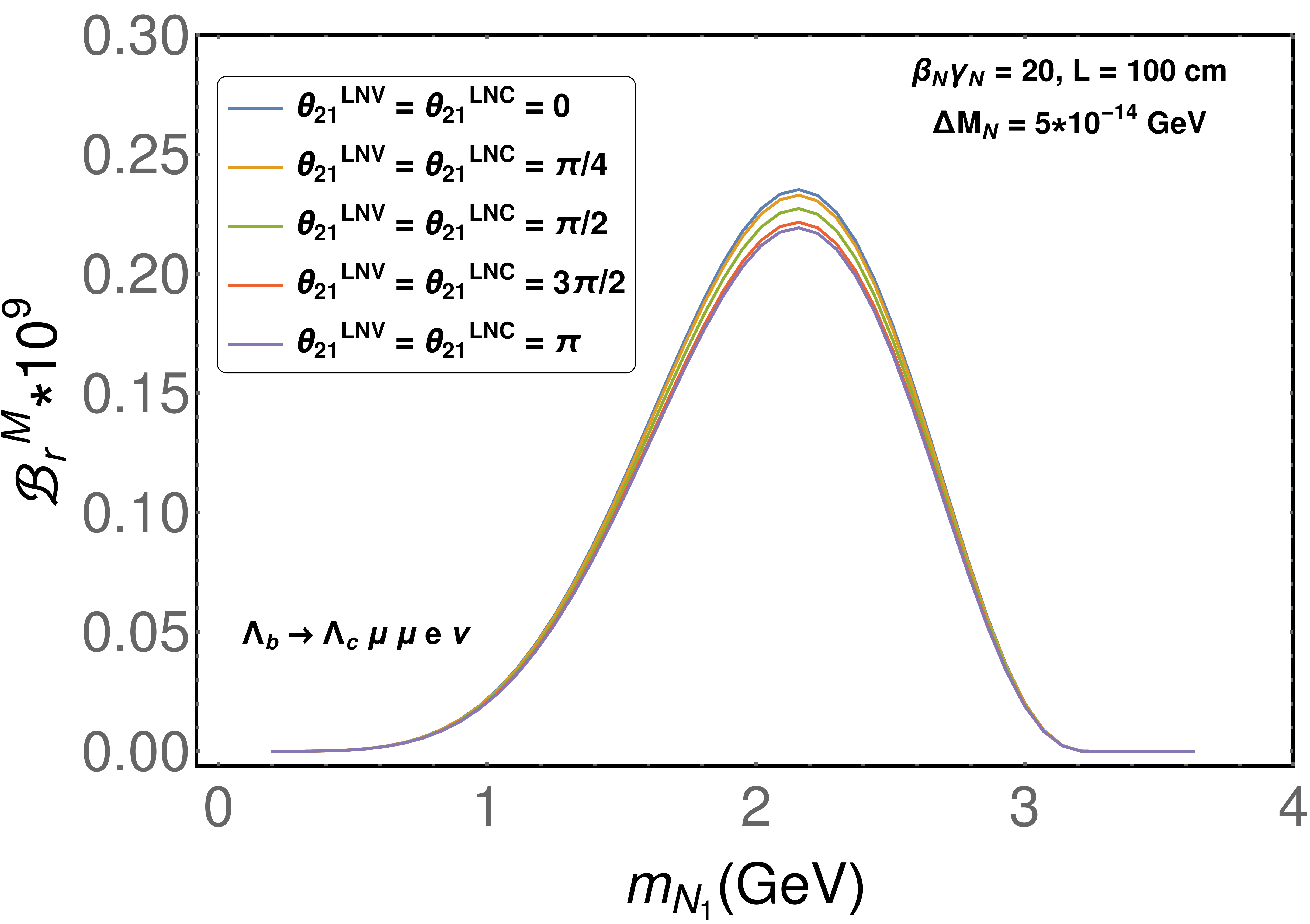}
		\includegraphics[scale=0.15]{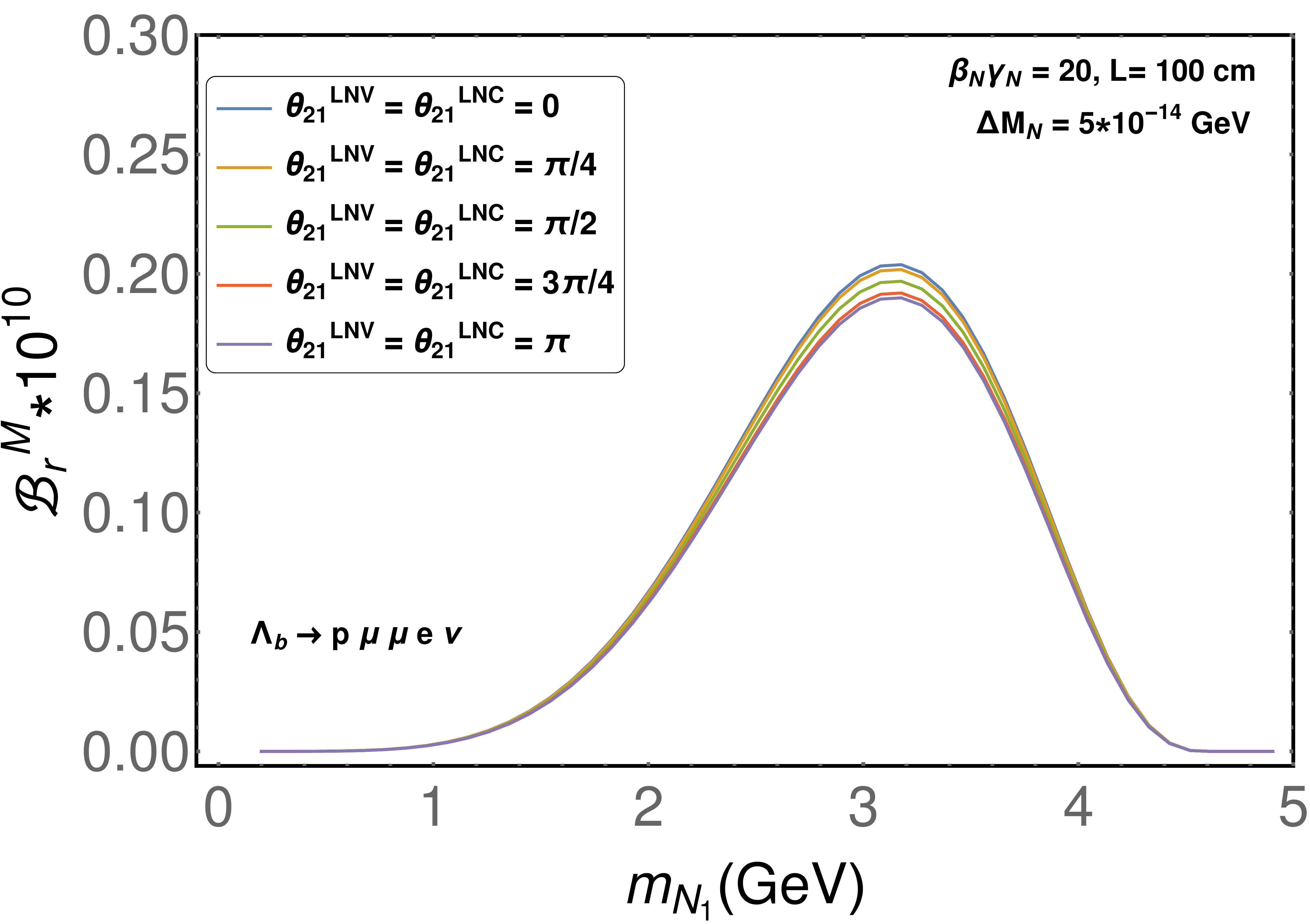}
		\caption{Branching ratios $\mathcal{B}r^{\rm M}(\Lambda_b^0\to \Lambda_c\mu\mu e\nu)$ and $\mathcal{B}r^{\rm M}(\Lambda_b^0\to p\mu\mu e\nu)$ as a function of $m_{N_1}$ for $|V_{\mu N_1}|^2 = |V_{\mu N_2}|^2 = 10^{-5} $, $|V_{e N_1}|^2 = |V_{e N_2}|^2 = 10^{-7}$, $\Delta M_N = 5\times10^{-14}$ GeV, $\beta_N \gamma_N = 20$, maximal displaced vertex length $L$ = 100 cm, $\tau_{N_1}=\tau_{N_2}=100 $ ps, and for different values of weak phase angles $\theta_{21}^{\rm LNV}$ and $\theta_{21}^{\rm LNC}$ \label{fig:brmNtheta} }
	\end{center}
\end{figure}

\begin{figure}[h!]
	\begin{center}
		\includegraphics[scale=0.15]{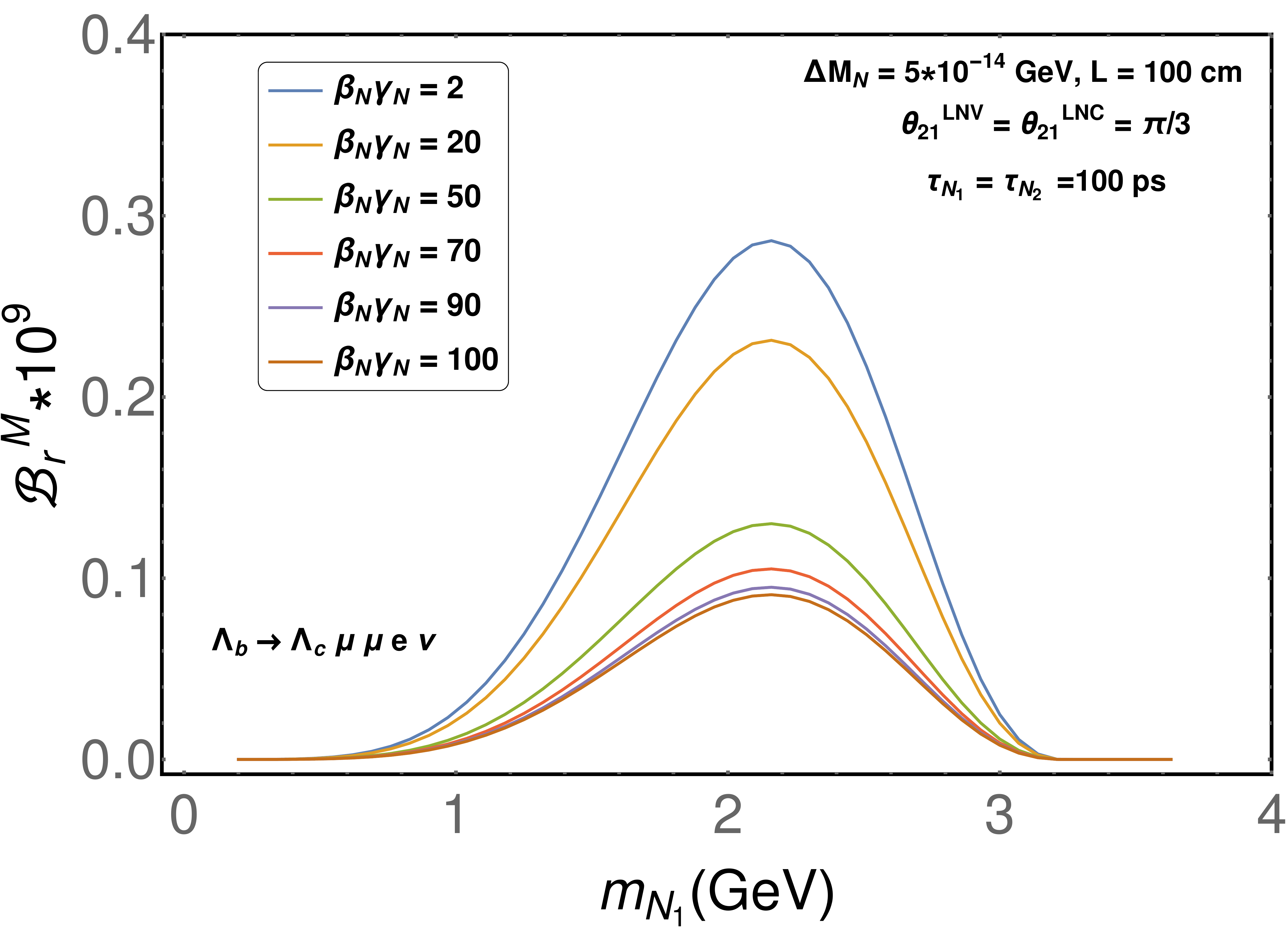}
		\includegraphics[scale=0.15]{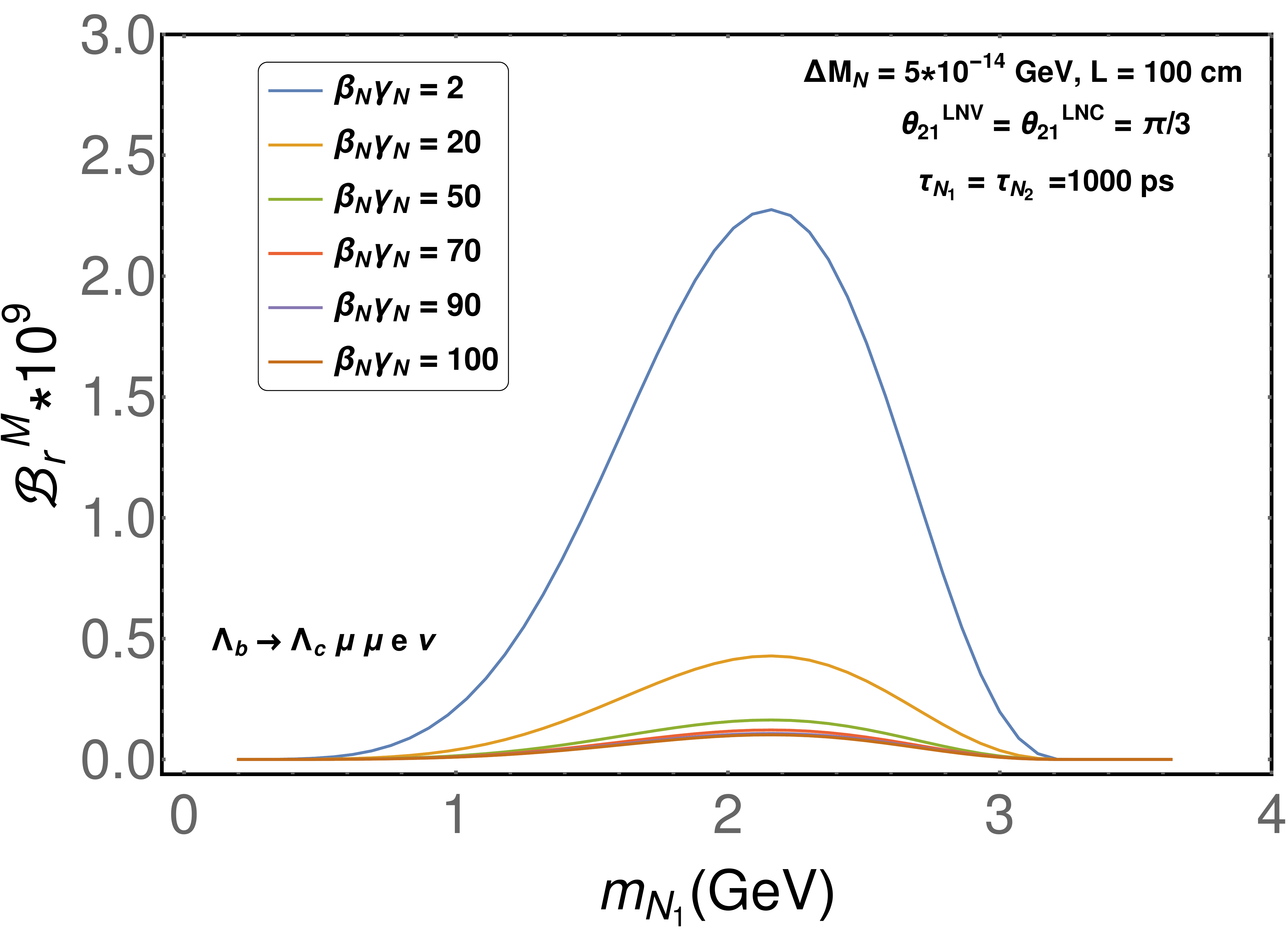}
		\includegraphics[scale=0.15]{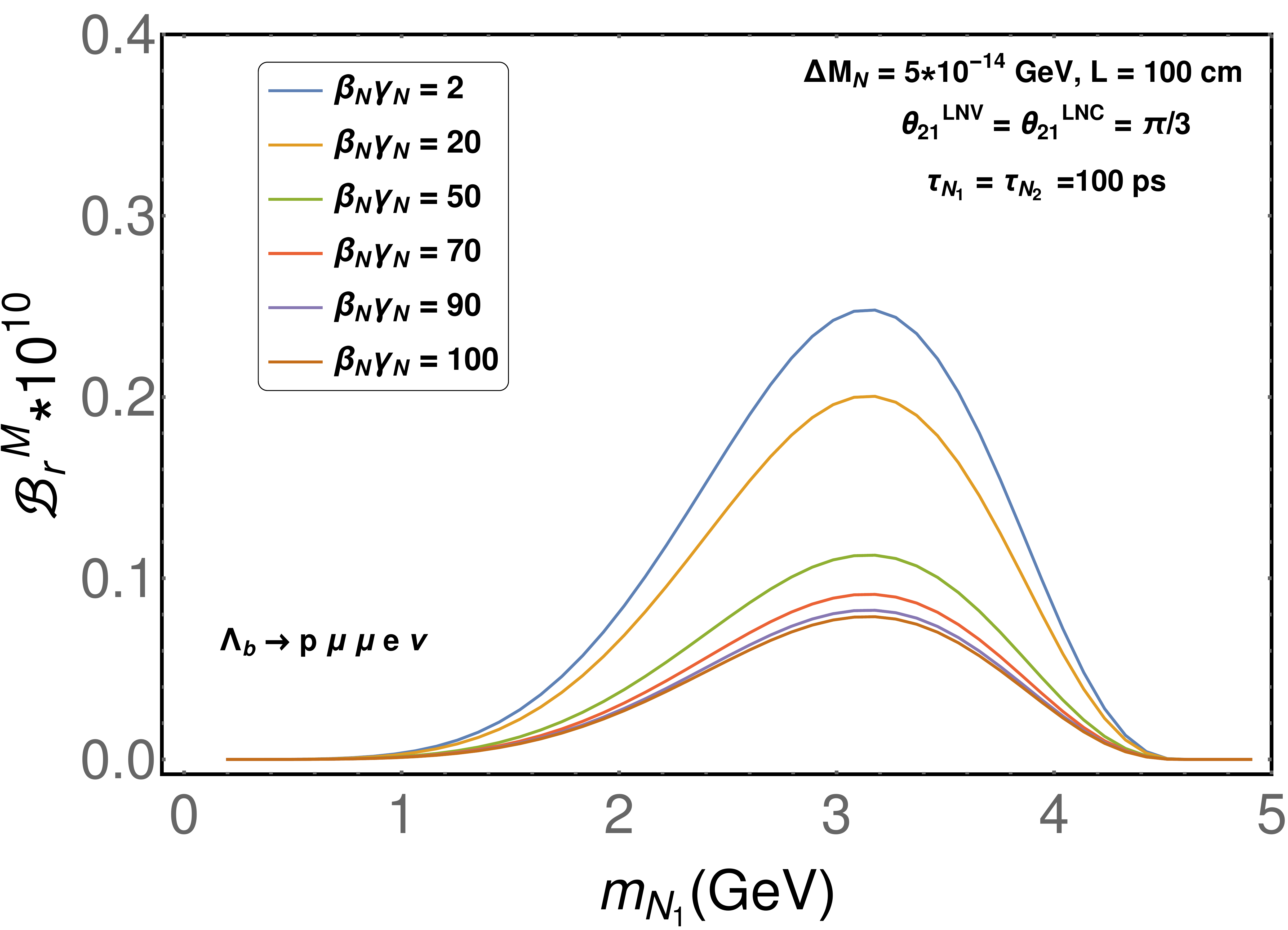}
		\includegraphics[scale=0.15]{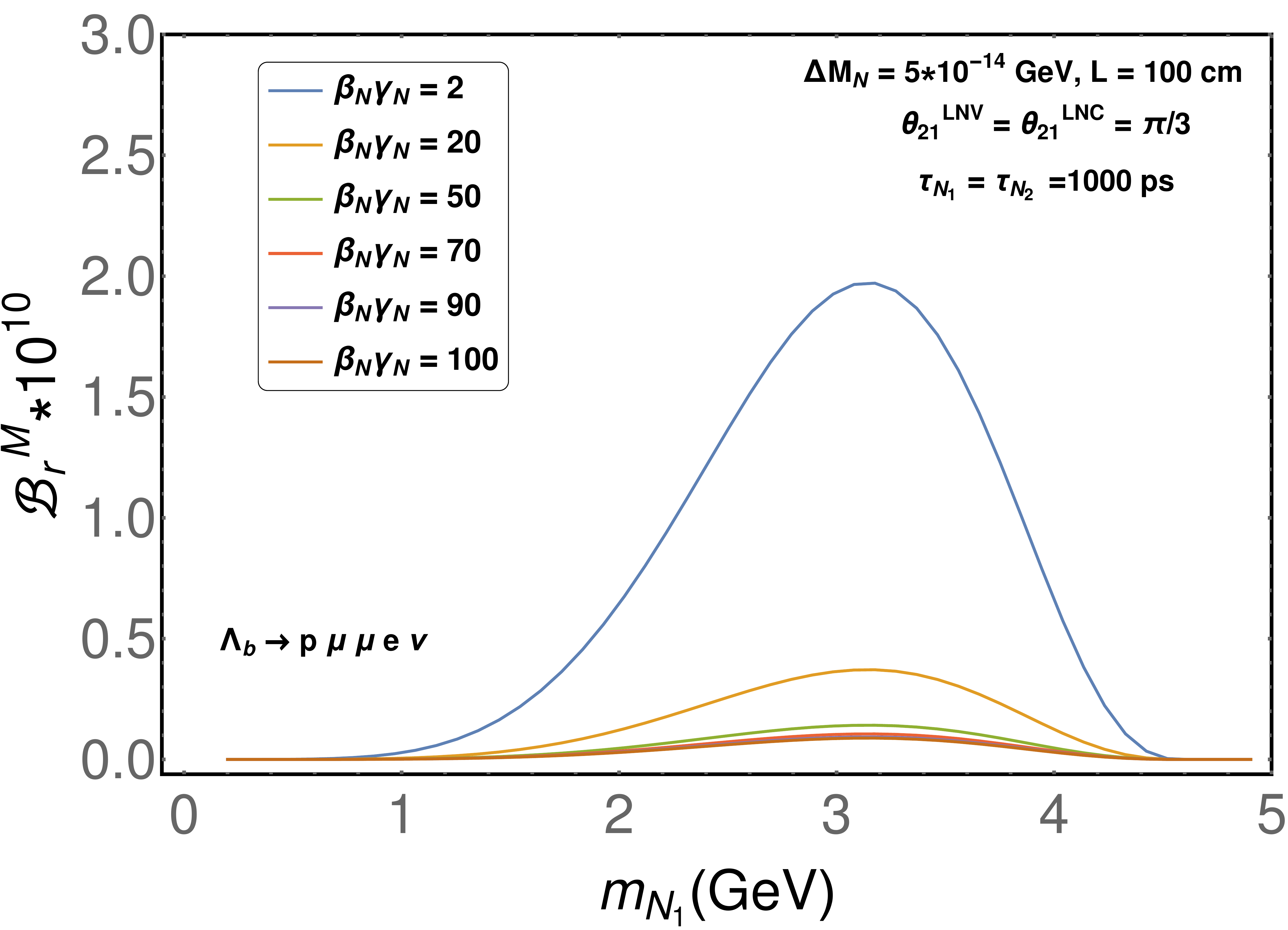}
		\caption{Branching ratios $\mathcal{B}r^{\rm M}(\Lambda_b^0\to (\Lambda_c,p)\mu\mu e\nu)$  as a function of $m_{N_1}$ for the values of heavy light mixing elements $|V_{\mu N_1}|^2 = |V_{\mu N_2}|^2 = 10^{-5} $, $|V_{e N_1}|^2 = |V_{e N_2}|^2 = 10^{-7}$ and $\Delta M_N = 5\times10^{-14}$ GeV, weak phase angles $\theta_{21}^{\rm LNV}= \theta_{21}^{\rm LNC}= \pi/3 $, neutrino flight length $L$= 100 cm and $\tau_{N_1}=\tau_{N_2}=[100,1000] $ ps for different values of Lorentz factors $\beta_N \gamma_N $. \label{fig:brmNbeta}}
	\end{center}
\end{figure}

\begin{figure}[h!]
	\begin{center}
		\includegraphics[scale=0.15]{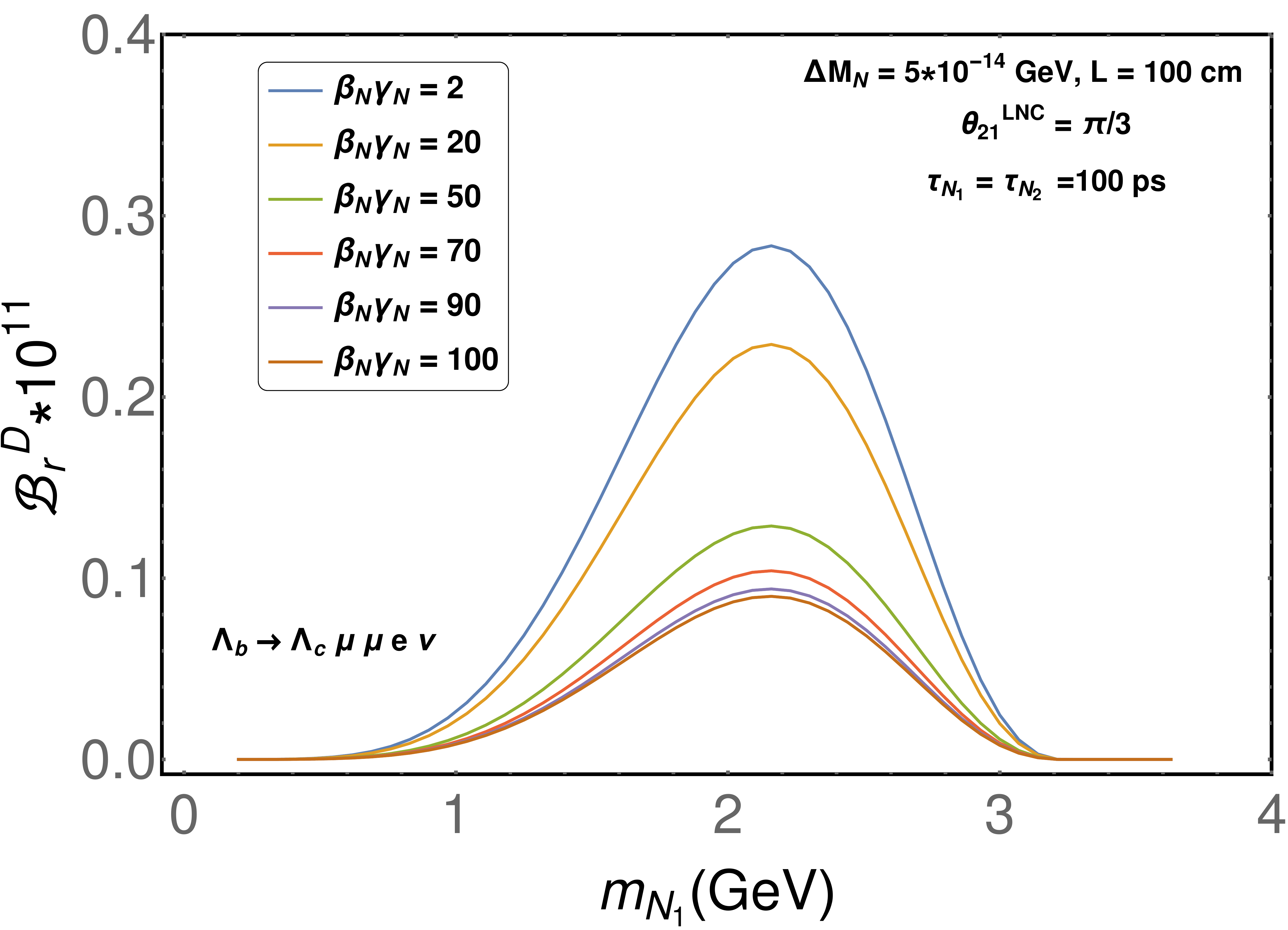}
		\includegraphics[scale=0.15]{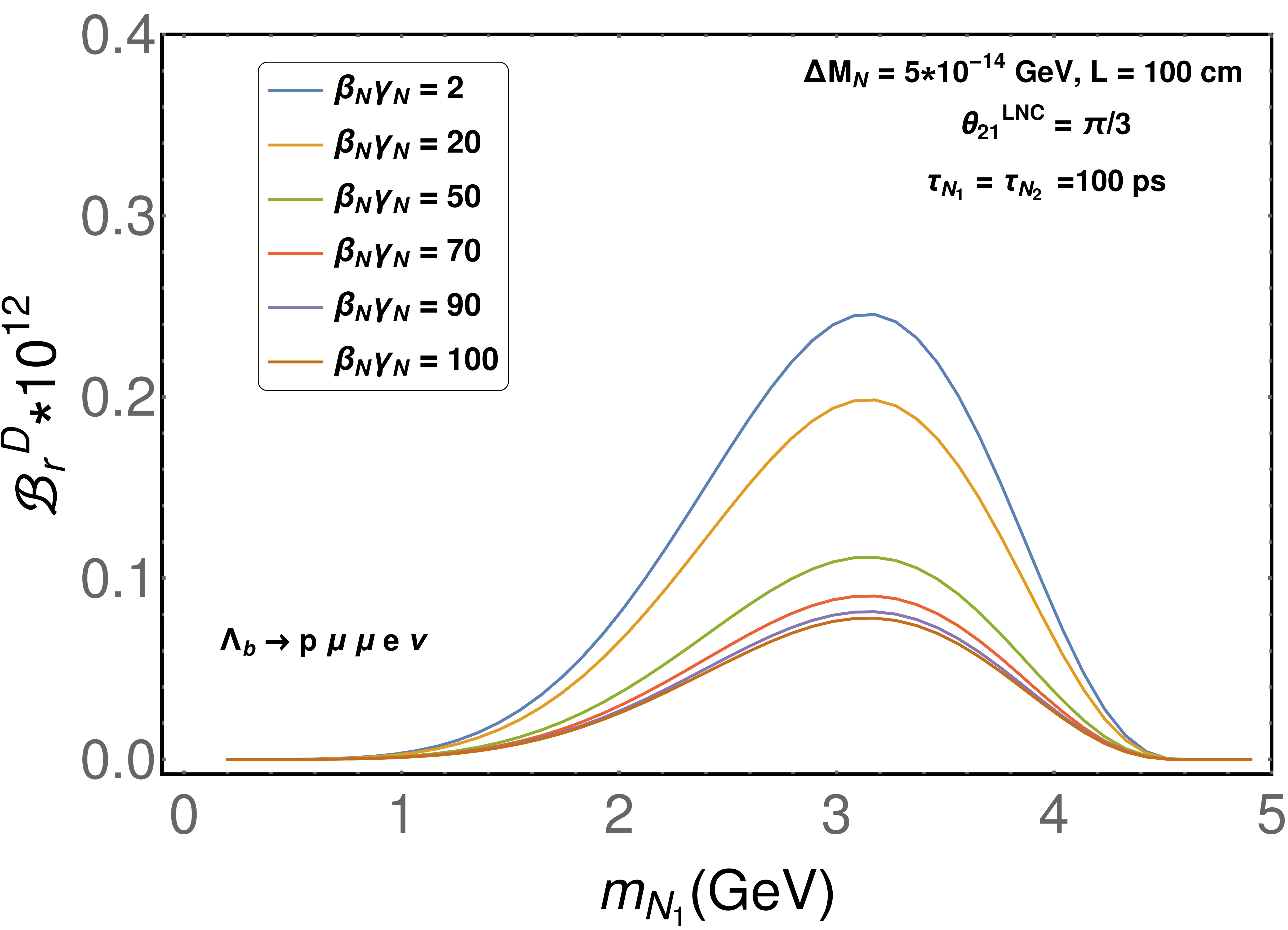}
		\caption{CP average branching ratios $\mathcal{B}r^{\rm D}(\Lambda_b^0\to (\Lambda_c,p)\mu\mu e\nu) $ mediated by Dirac neutrinos  as a function of $m_{N_1}$ for the values of heavy light mixing elements $|V_{\mu N_1}|^2 = |V_{\mu N_2}|^2 = 10^{-5} $, $|V_{e N_1}|^2 = |V_{e N_2}|^2 = 10^{-7}$ and $\Delta M_N = 5\times 10^{-14}$ GeV, weak phase angles $\ \theta_{21}^{\rm LNC}= \pi/3 $, neutrino flight length $L$= 100 cm and $\tau_{N_1}=\tau_{N_2}=100 $ ps for different values of Lorentz factors $\beta_N \gamma_N $. \label{fig:brmNbetaDirac}}
	\end{center}
\end{figure}
 
 \begin{figure}[h!]
	\begin{center}
		\includegraphics[scale=0.15]{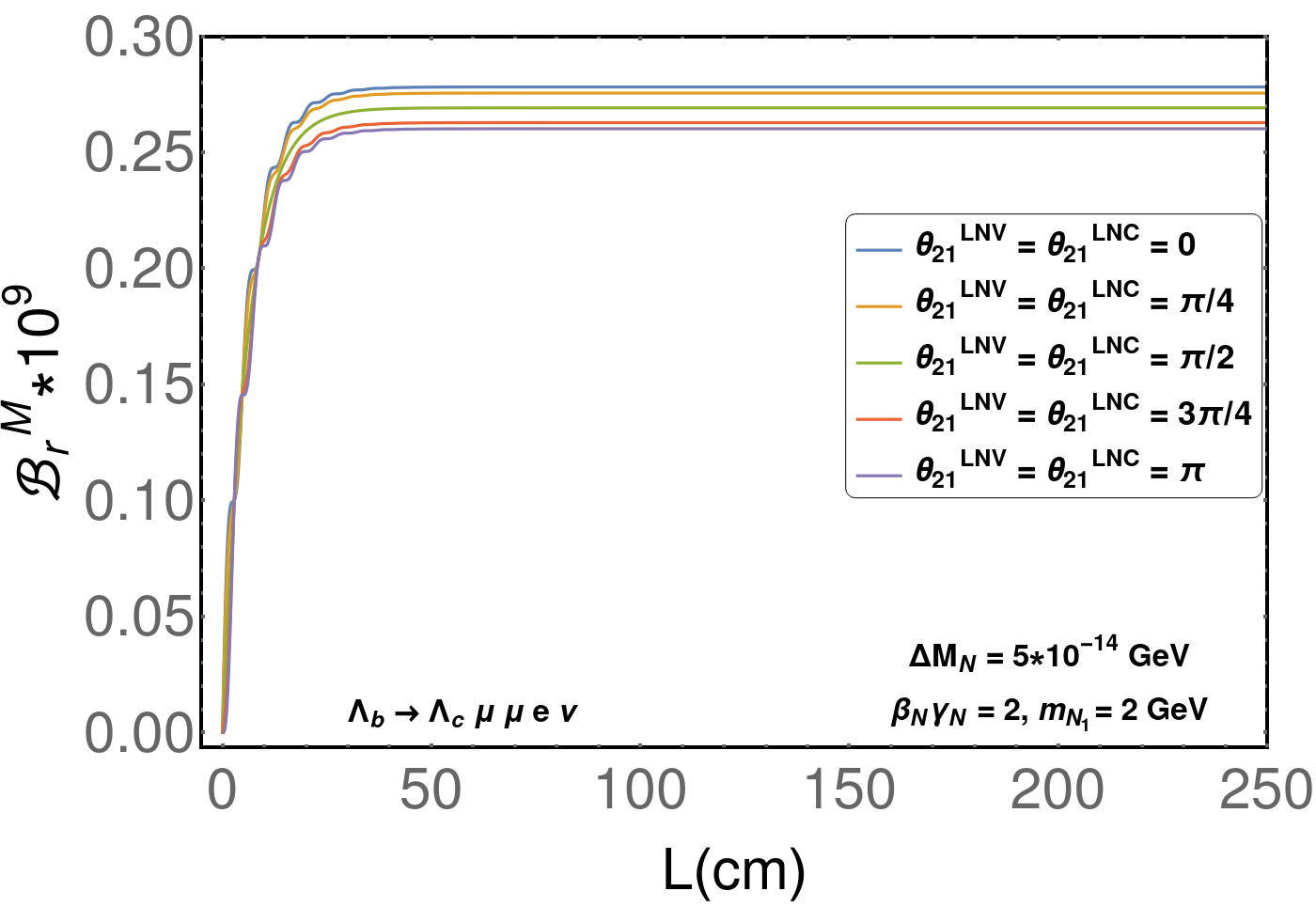}
		\includegraphics[scale=0.15]{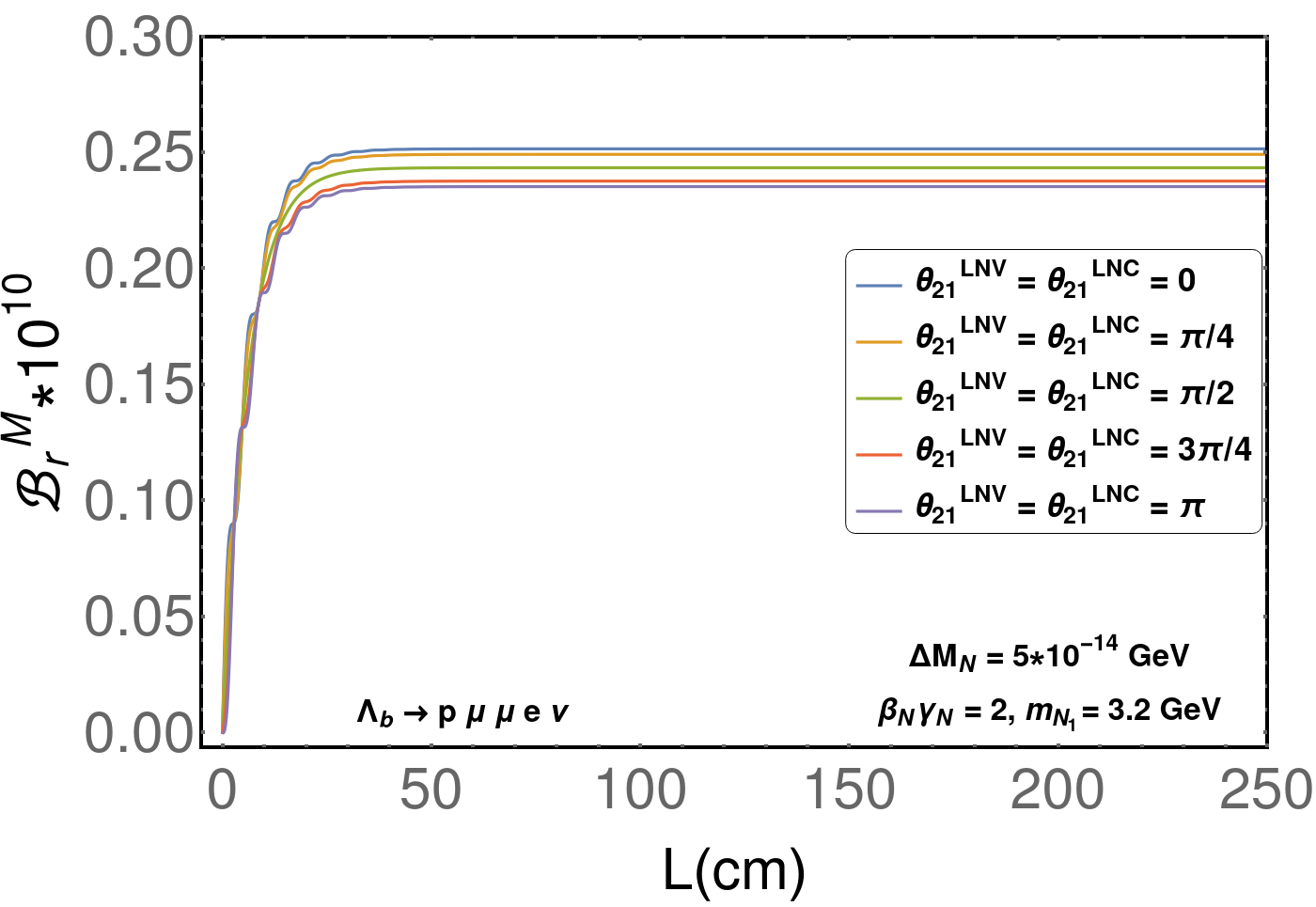}
		\includegraphics[scale=0.15]{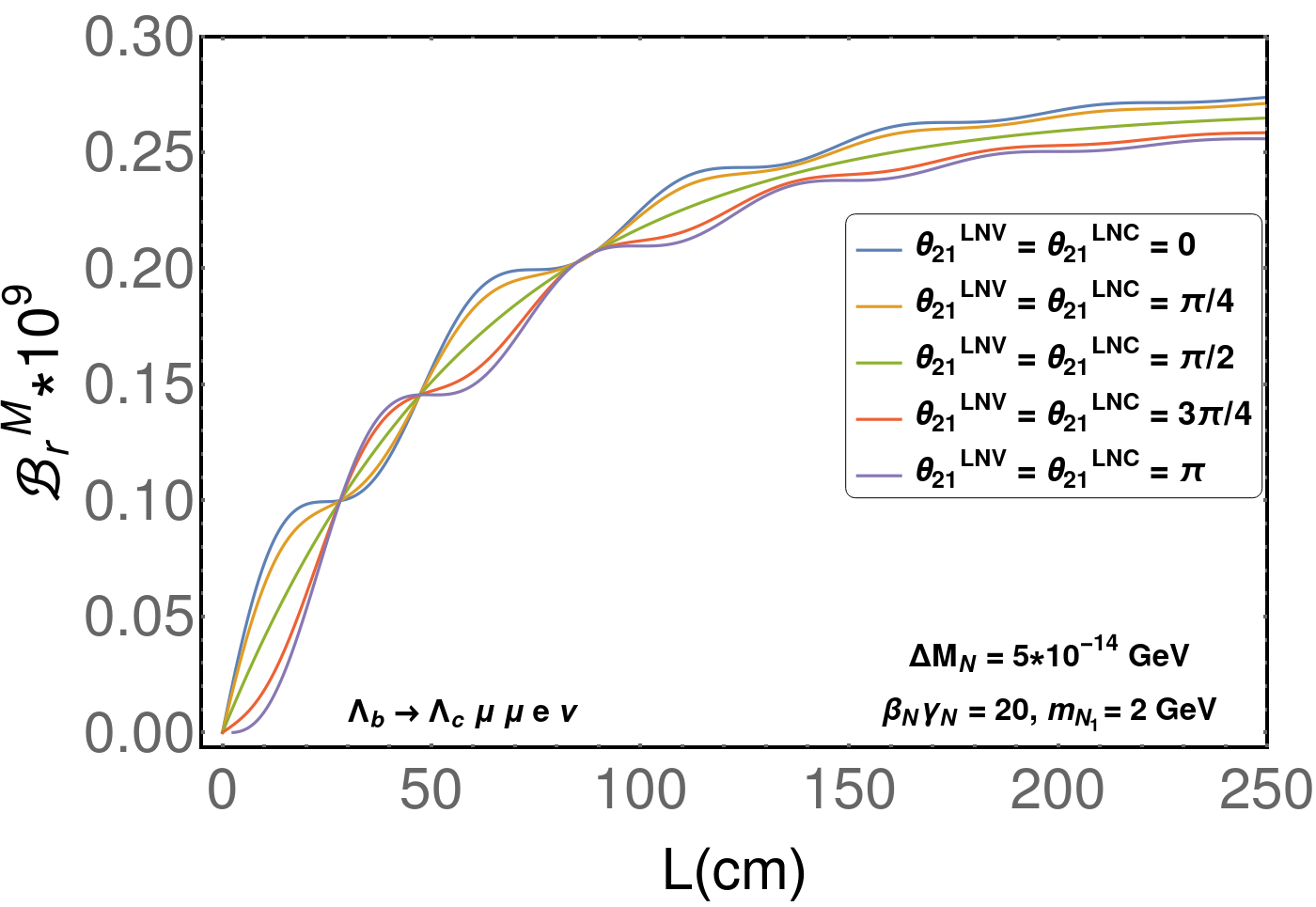}
		\includegraphics[scale=0.15]{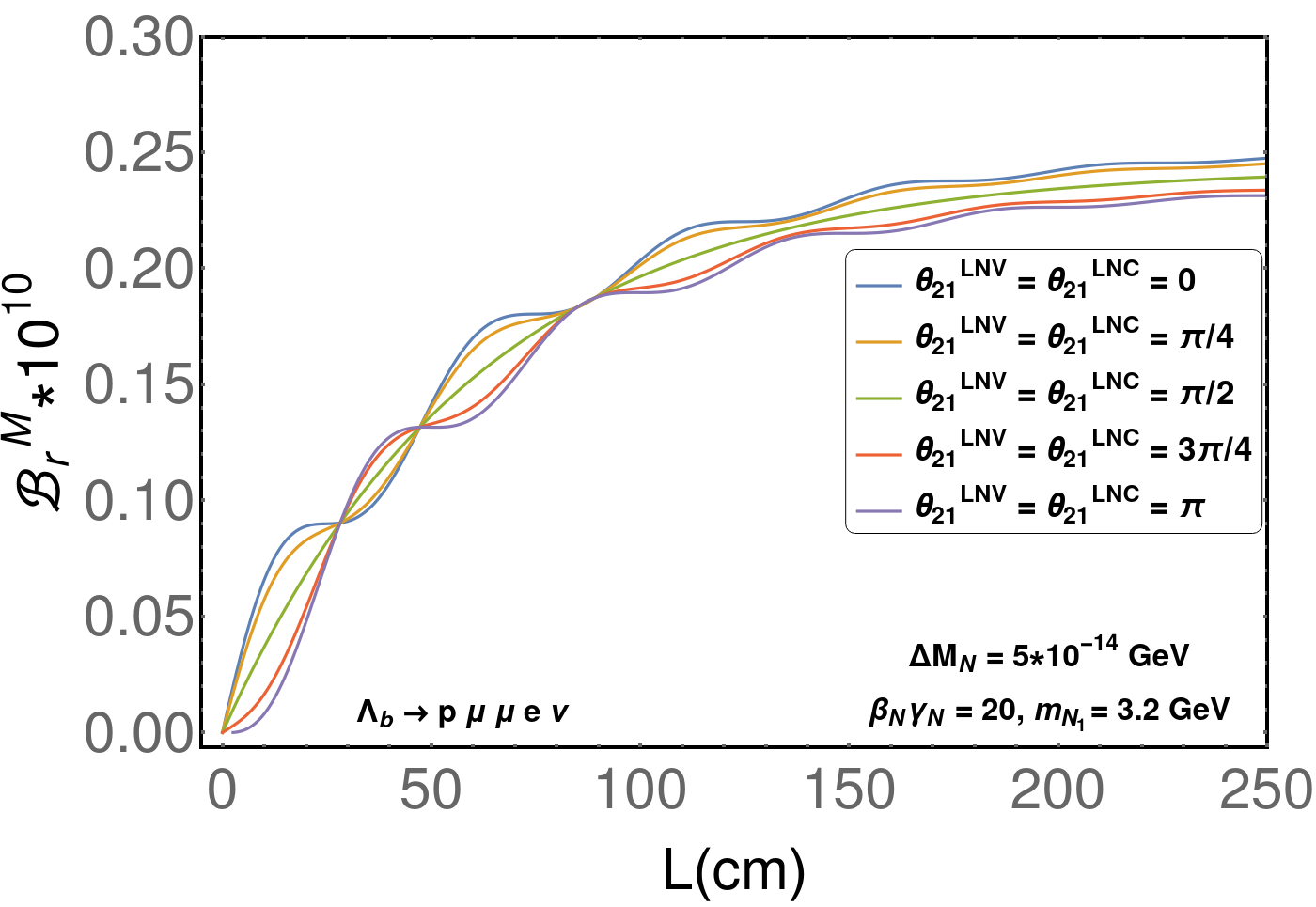}
		\caption{Branching ratios $\mathcal{B}r^{\rm M}(\Lambda_b^0\to (\Lambda_c,p)\mu\mu e\nu) $  as a function of the neutrino flight length $L$ in cm for different values of weak phase angles $\theta_{21}^{\rm LNV}$, $ \theta_{21}^{\rm LNC}$ and for the values of heavy light mixing $|V_{\mu N_1}|^2 = |V_{\mu N_2}|^2 = 10^{-5} $,  $|V_{e N_1}|^2 = |V_{e N_2}|^2 = 10^{-7}$ and mass difference $\Delta M_N = 5 \times 10^{-14} $ GeV, Lorentz factors $\beta_N \gamma_N $= [2,20], Majorana mass $M_{N_1}$ = 2 GeV and 3.2 GeV for $\Lambda_c$ and proton receptively, $\tau_{N_1}=\tau_{N_2}=100 $ ps.   \label{fig:brL}}
	\end{center}
\end{figure}

\begin{figure}[h!]
	\begin{center}
		\includegraphics[scale=0.15]{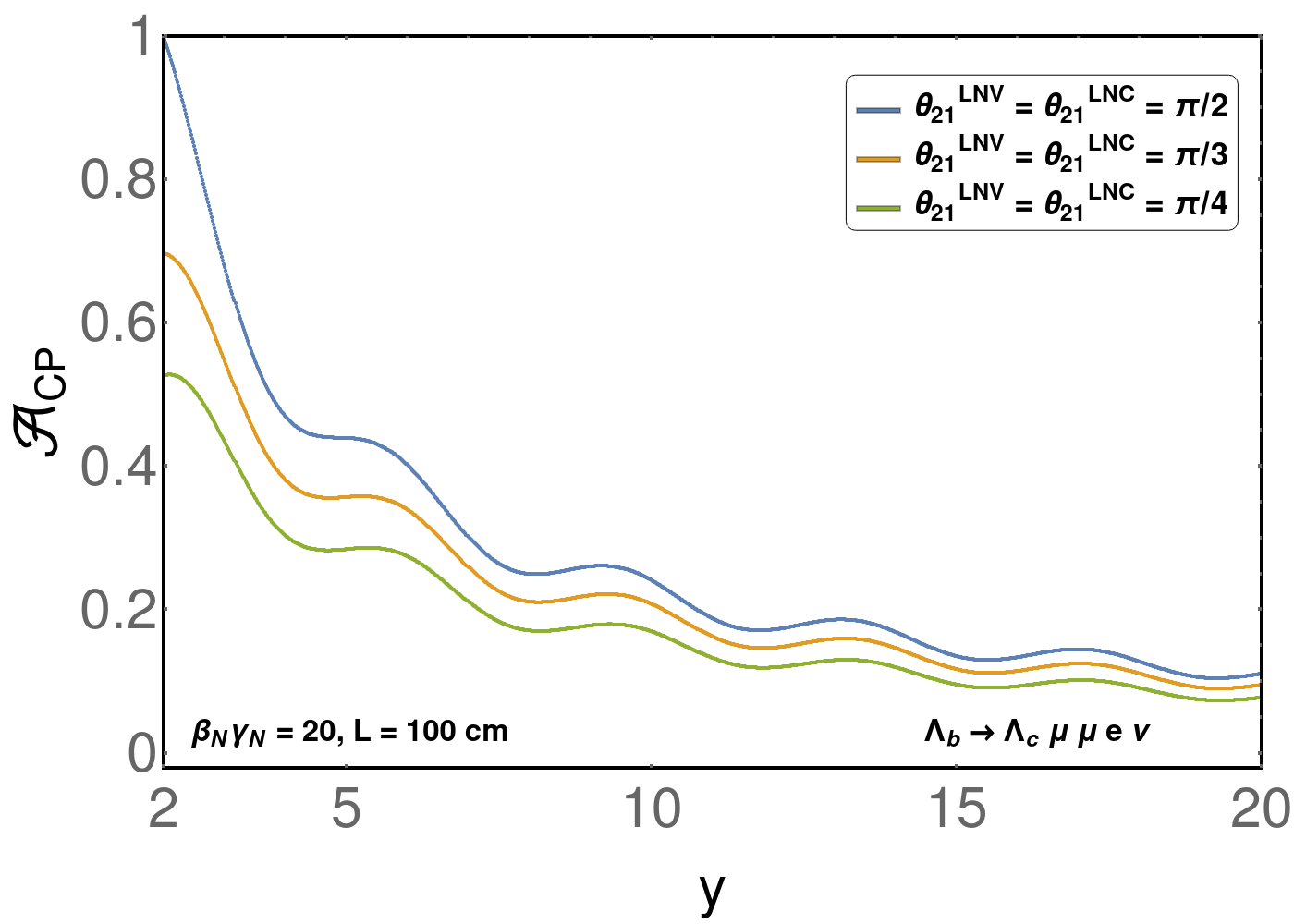}
		\includegraphics[scale=0.15]{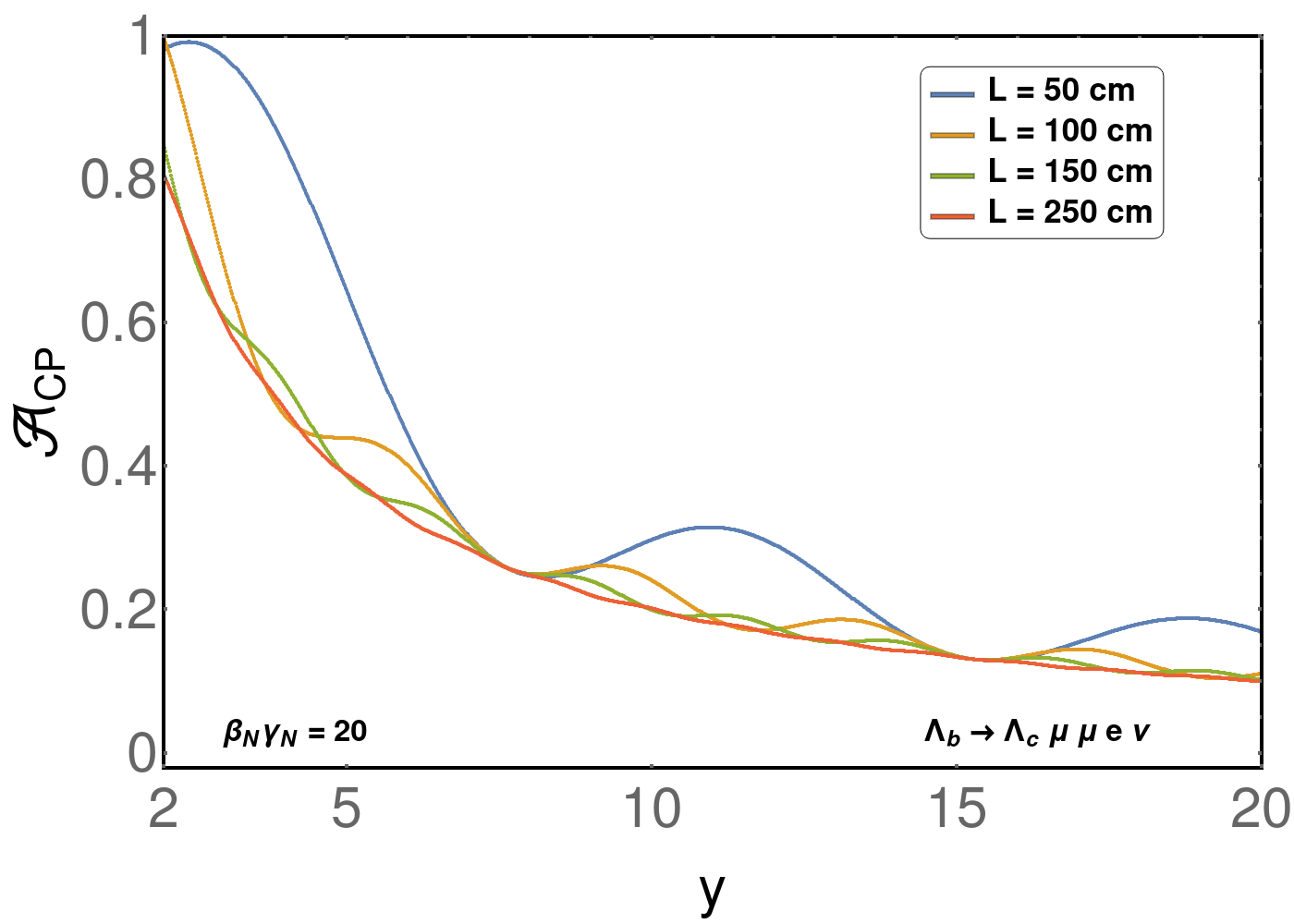}
		\caption{The CP asymmetry observable $\mathcal{A}_{CP}$ for $\Lambda_b^0 \to \Lambda_c \mu \mu e \nu$ as a function of $y = \frac{\Delta M_N}{\Gamma_N}$ for different values of phases (left panel) and maximal decay vertex (right panel) and for the values of $|V_{\mu N_1}|^2= |V_{\mu N_2}|^2= 10^{-5}$, $|V_{e N_1}|^2= |V_{e N_2}|^2= 10^{-7}$ and $\tau_{N} = 100$ ps. Similar plots are obtained for $\Lambda_b^0 \to p \mu \mu e \nu$ and $\Lambda_b^0 \to (\Lambda_c,p) e e \mu \nu$.\label{fig:acpYthetea}}
	\end{center}
\end{figure}

\begin{figure}[h!]
	\begin{center}
		\includegraphics[scale=0.15]{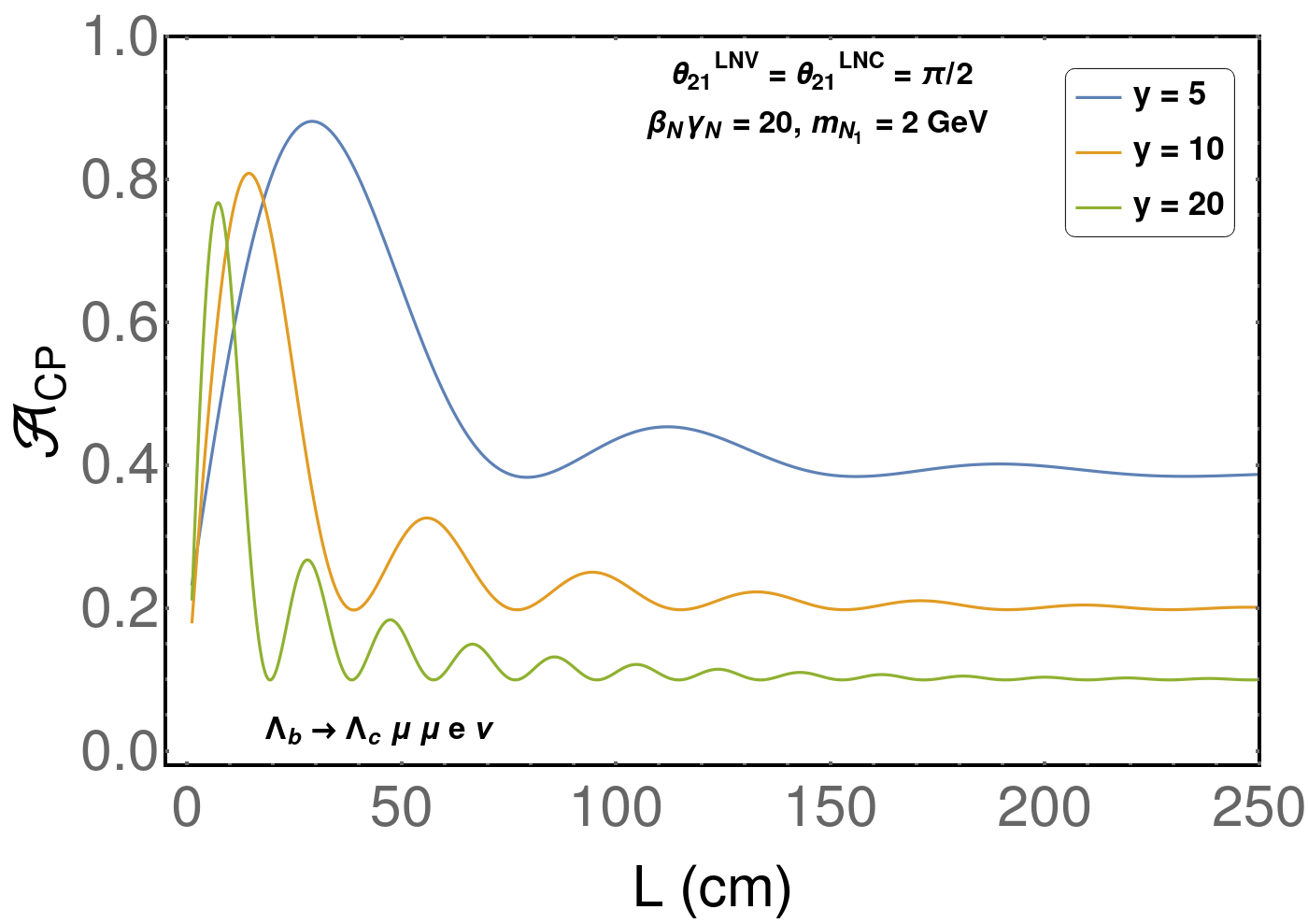}
		\includegraphics[scale=0.15]{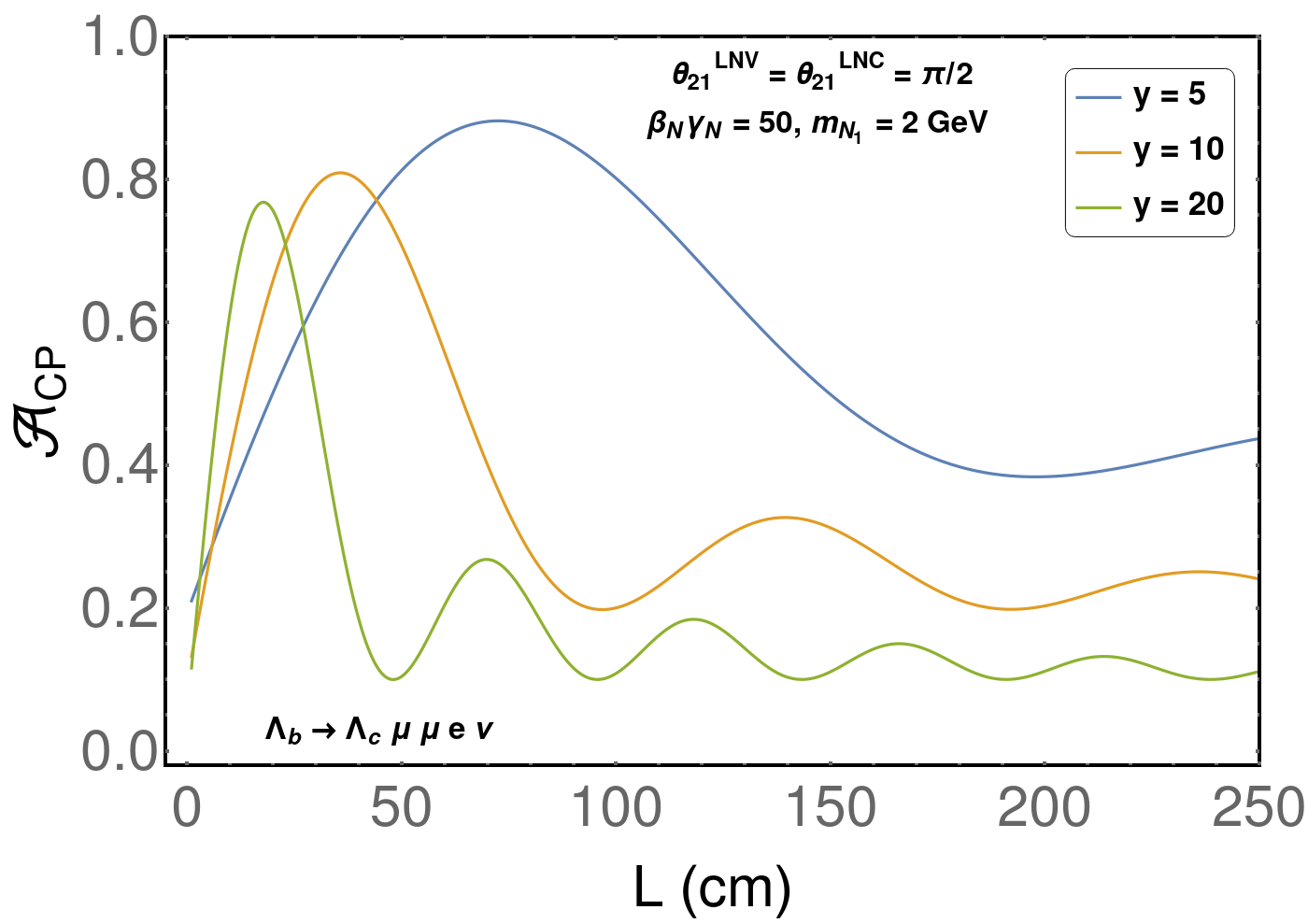}
		\caption{The CP asymmetry observable $\mathcal{A}_{CP}$ as a function of displaced vertex length $L$ for the decay mode $\Lambda_b^0 \to \Lambda_c \mu \mu e \nu$ for different values of $y = \frac{\Delta M_N}{\Gamma_N}$ and for the values of  $|V_{\mu N_1}|^2= |V_{\mu N_2}|^2= 10^{-5}$, $|V_{e N_1}|^2= |V_{e N_2}|^2= 10^{-7}$, $M_{N1}= 2$ GeV, $\theta_{21}^{\rm LNV}$= $\theta_{21}^{\rm LNC}$ = $\pi/2$, $\beta_N\gamma_N$ = 20, $\tau_{N_1} = \tau_{N_2} = 100$ ps. An identical plots are obtained for $\Lambda_b^0 \to p \mu \mu e \nu$ and $\Lambda_b^0 \to (\Lambda_c,p) e e \mu \nu$. \label{fig:acpL}}
\end{center}
\end{figure}

\begin{figure}[h!]
	\begin{center}
	    \includegraphics[scale=0.15]{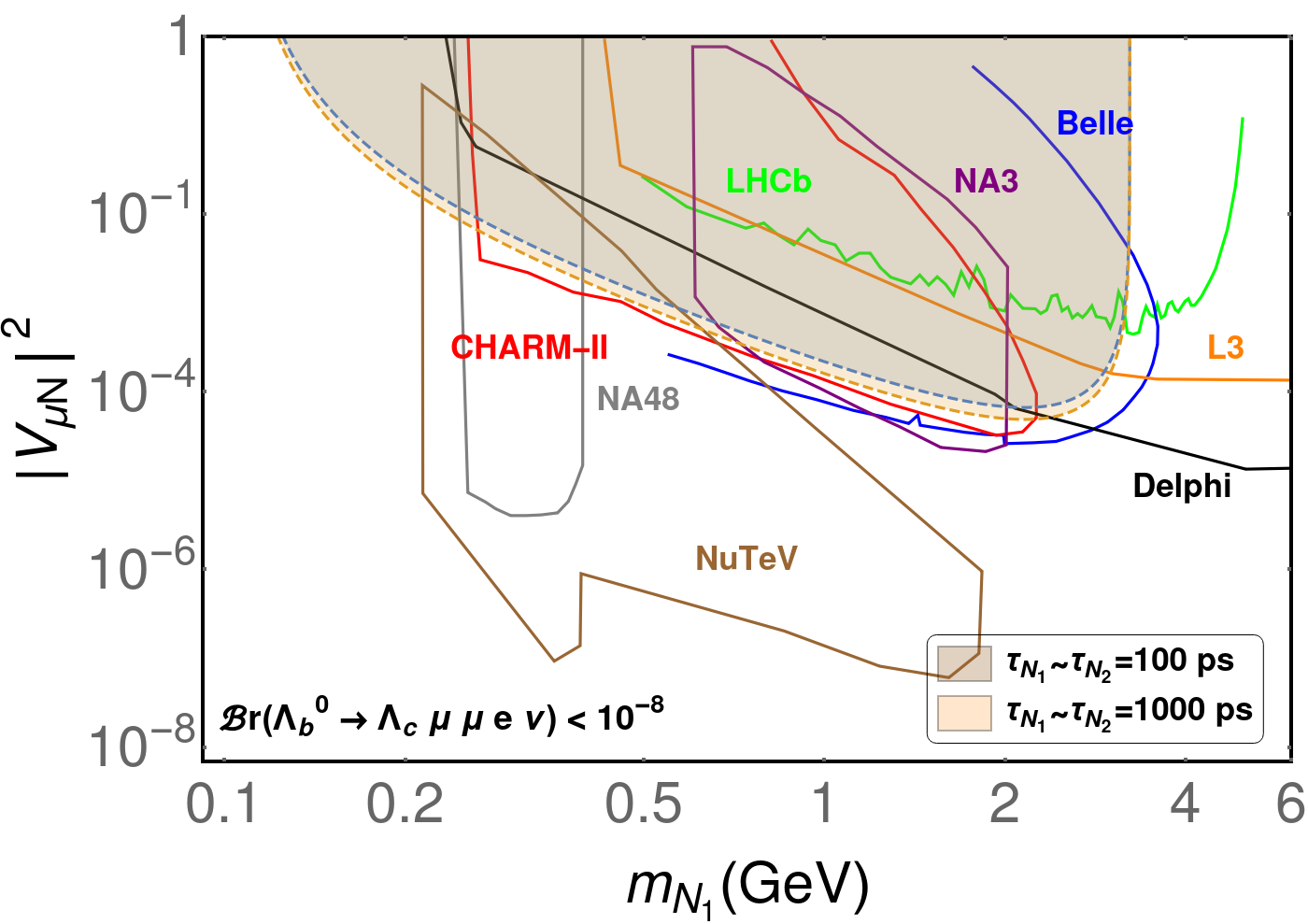}
	    \includegraphics[scale=0.15]{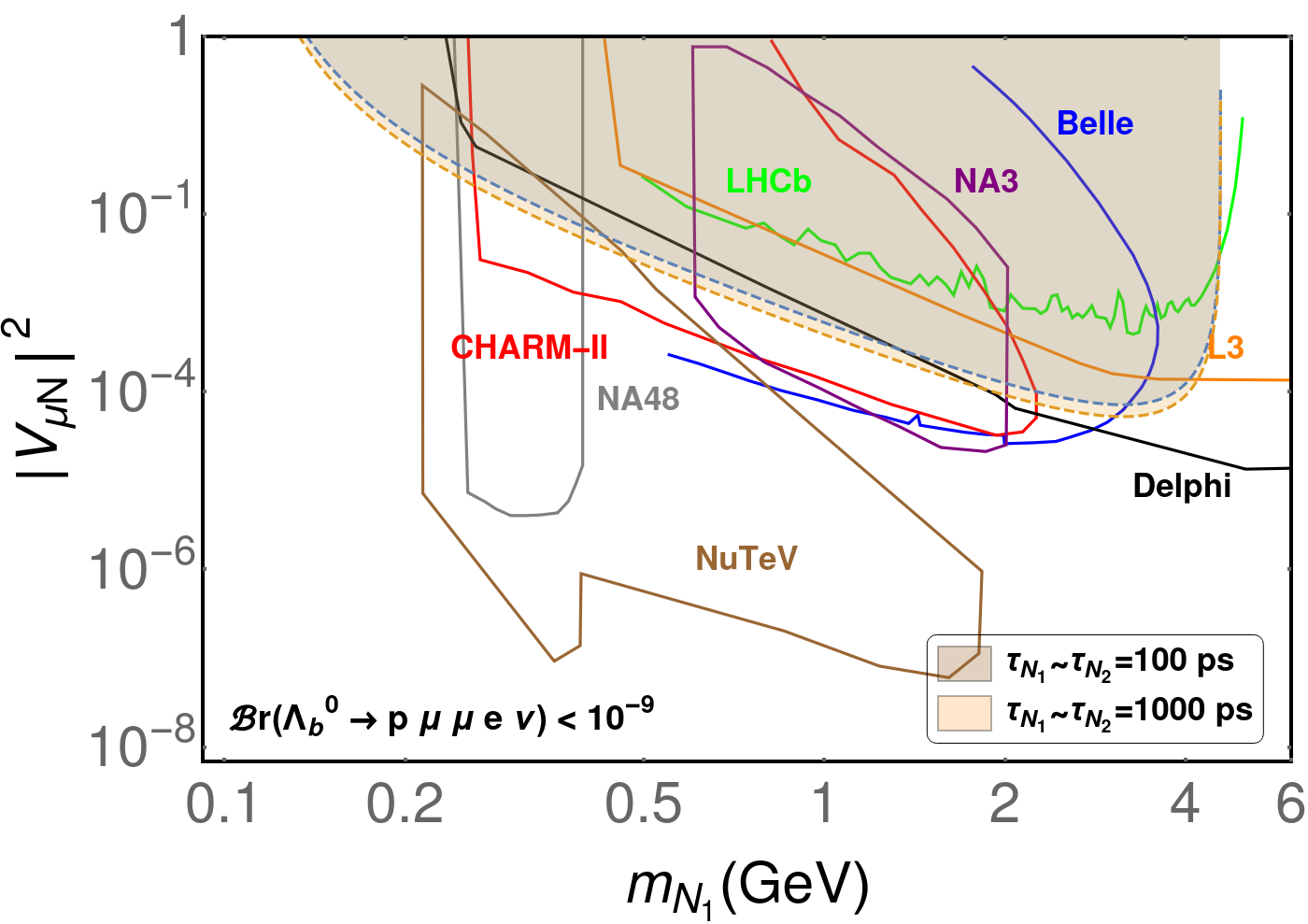}
		\includegraphics[scale=0.15]{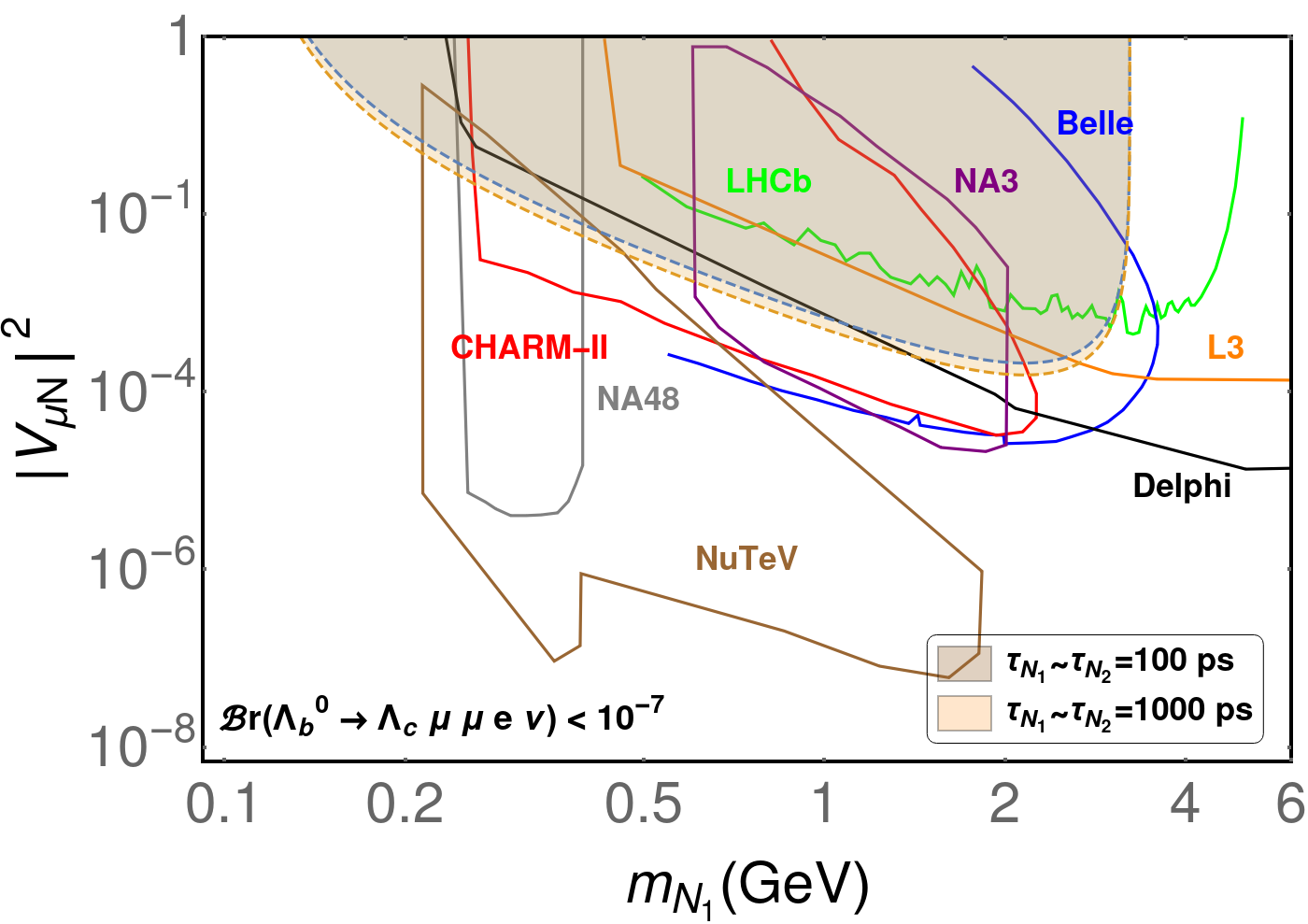}
		\includegraphics[scale=0.15]{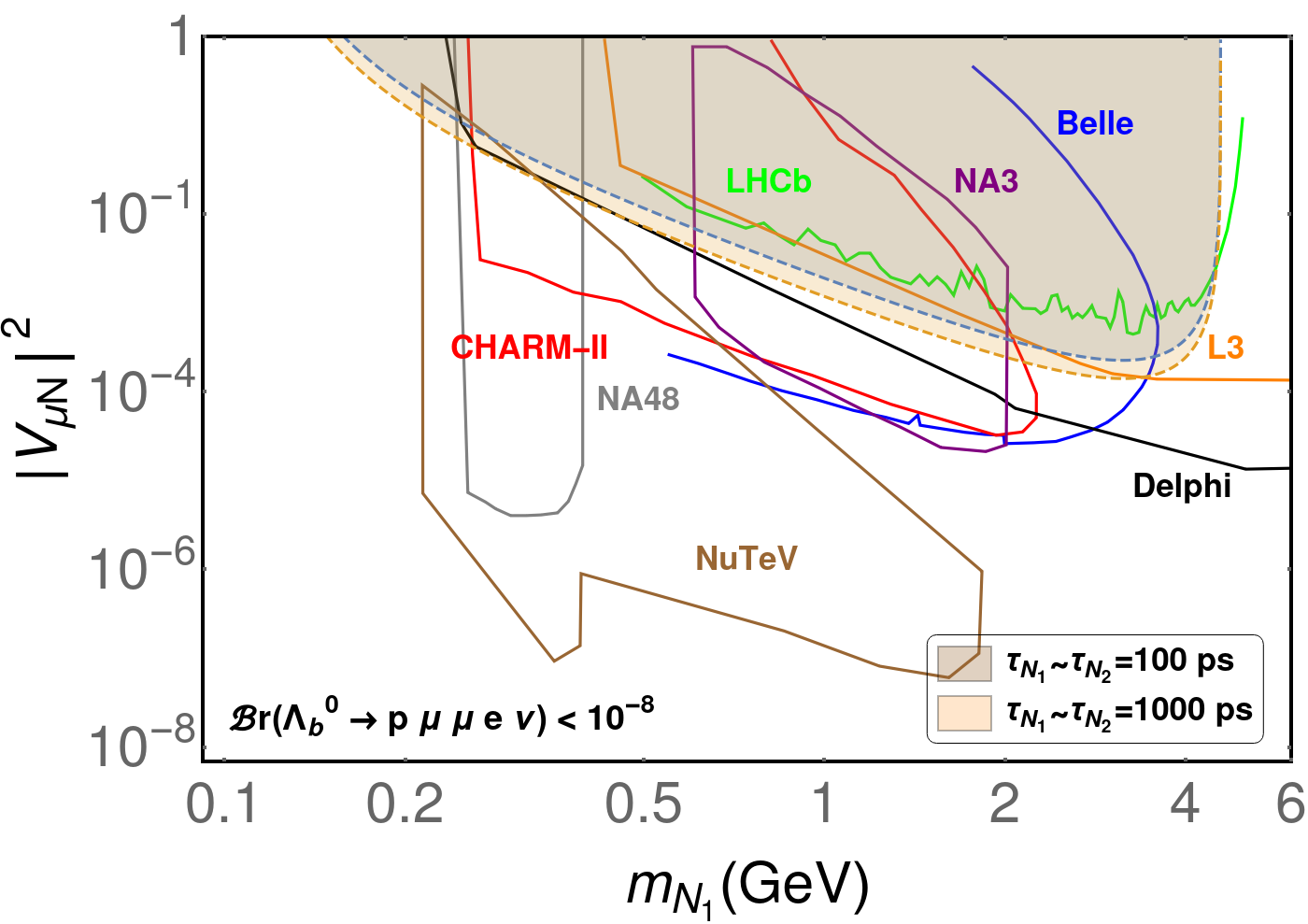}
		\caption{Exclusion regions on the ($|V_{\mu N}|^2, m_{N_1}$) parameter space for the case of LNV decays mediated by Majorana neutrinos for $\mathcal{B}r^{\rm M}(\Lambda_b^0 \to p \mu \mu e \nu) < 10^{-8}, 10^{-9} $ and $\mathcal{B}r^{\rm M}(\Lambda_b^0 \to \Lambda_c  \mu \mu e \nu) < 10^{-7}, 10^{-8} $ for the values of $\tau_{N_1} = \tau_{N_2} = [100,1000]$ ps, $\theta_{21}^{\rm LNV}= \pi/3$ , $\Delta M_N = 5\times 10^{-14} $ GeV, $\beta_N\gamma_N= 20$ and $L= 100$ cm. \label{fig:cons}}
\end{center}
\end{figure}

\begin{figure}[h!]
	\begin{center}
	    \includegraphics[scale=0.15]{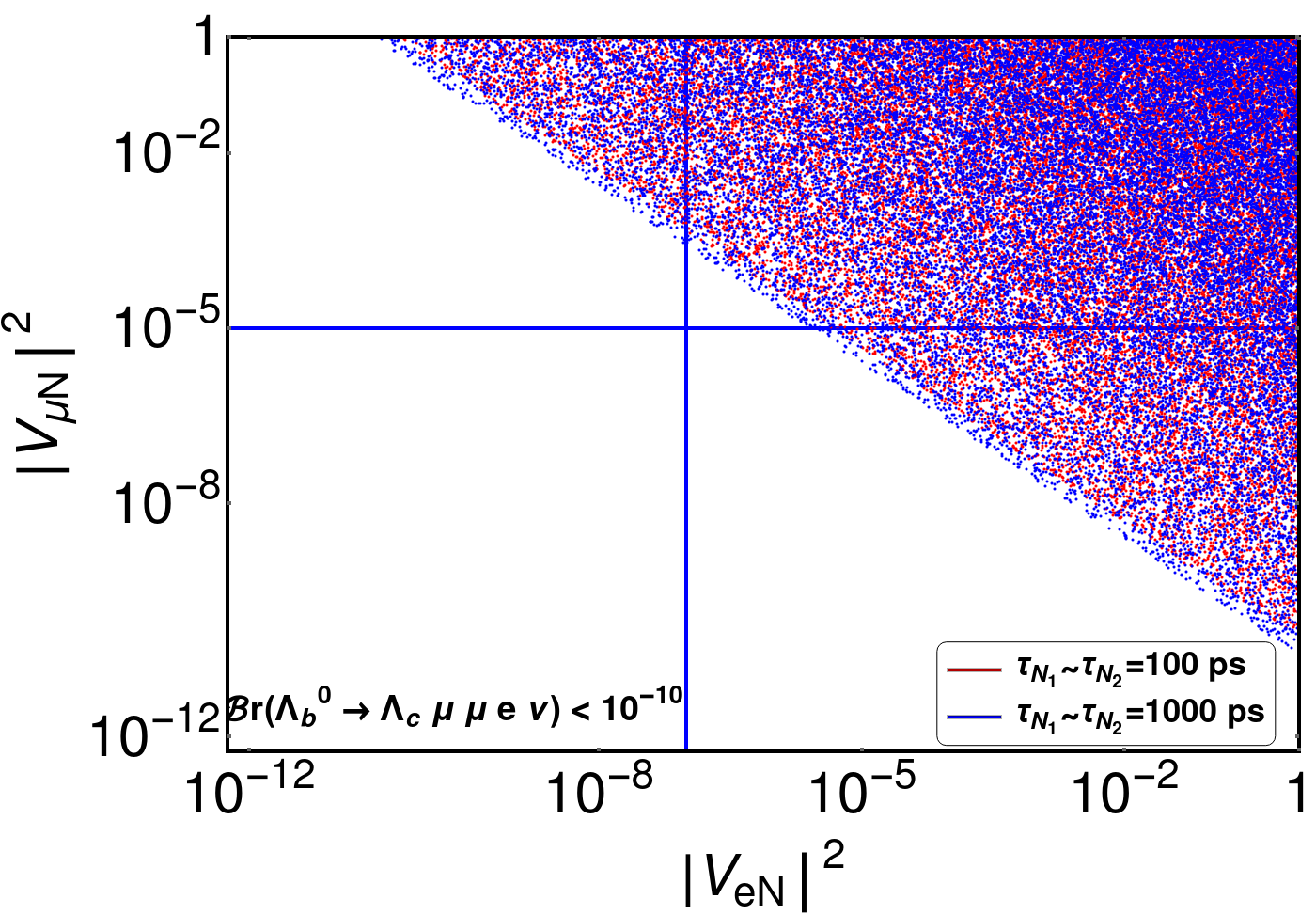}	    
		\includegraphics[scale=0.15]{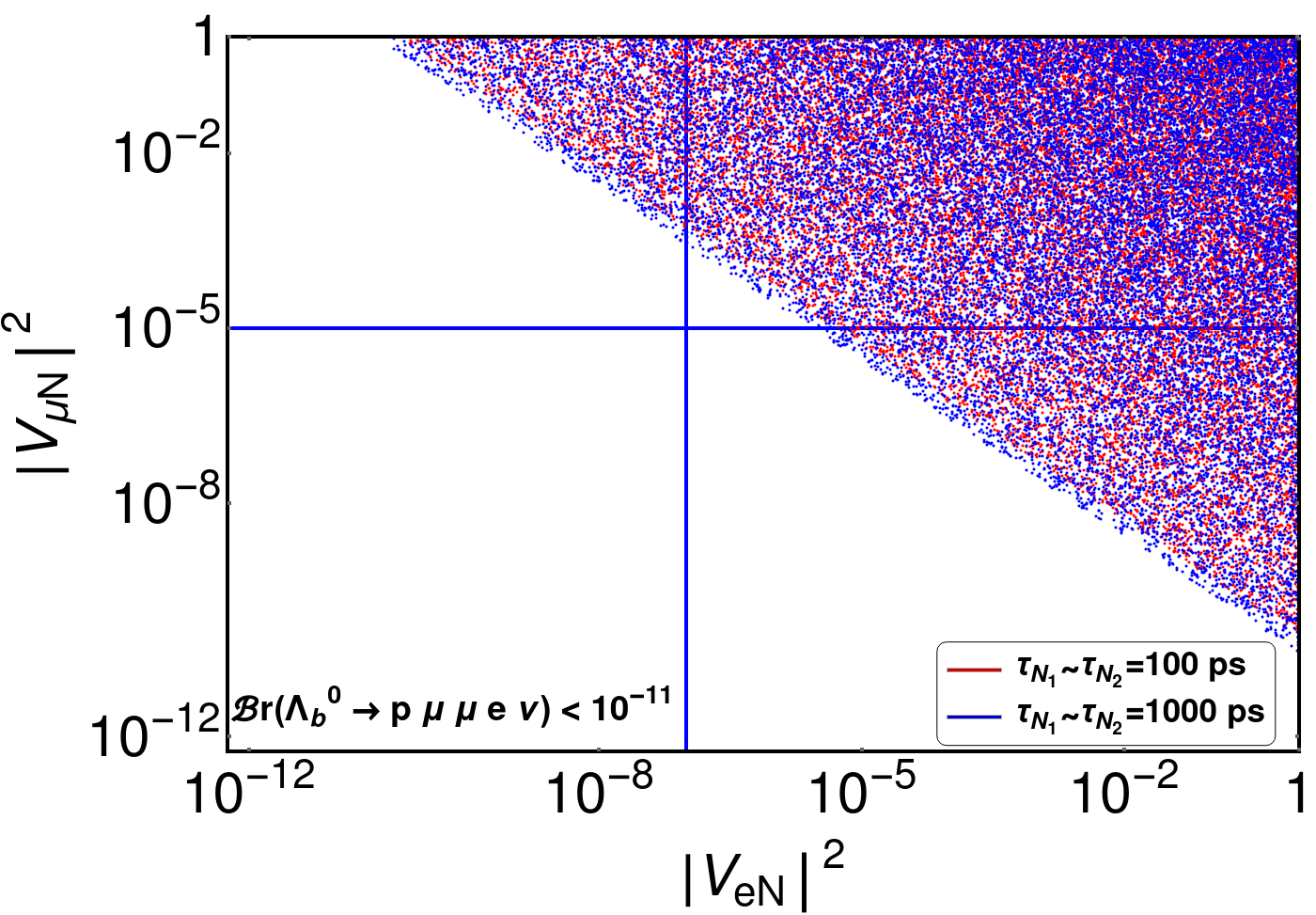}
		\includegraphics[scale=0.15]{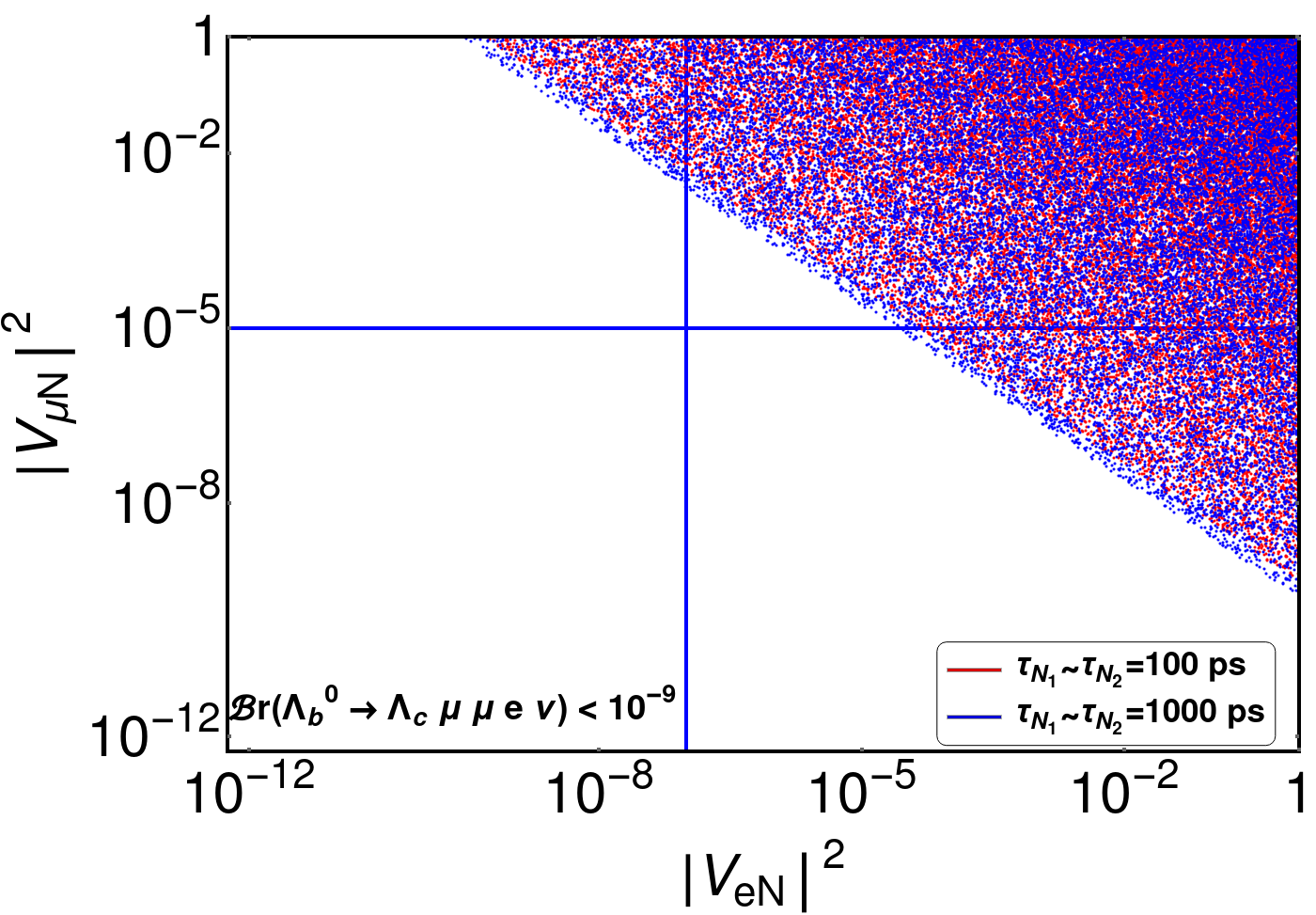}
		\includegraphics[scale=0.15]{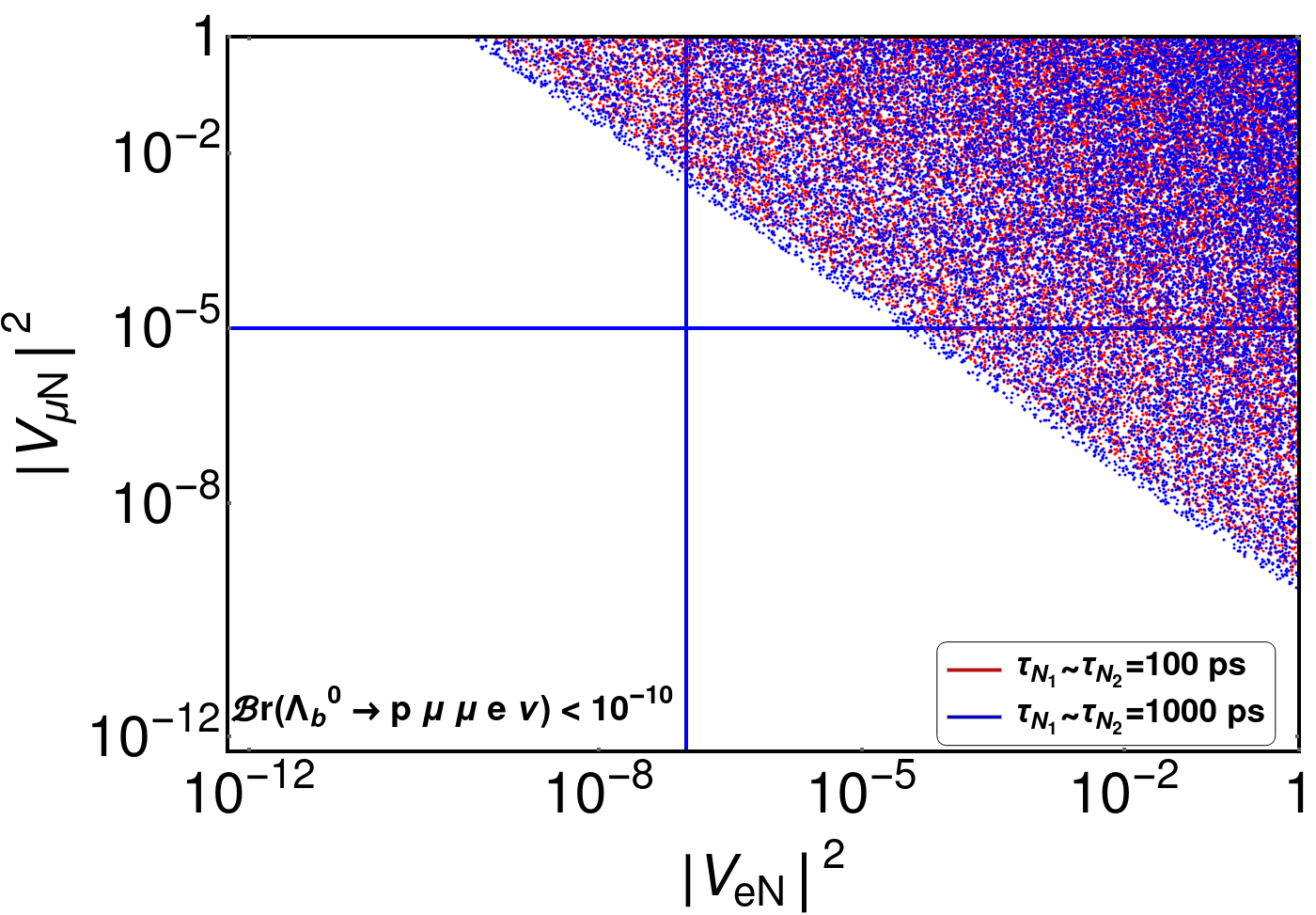}
		\caption{Exclusion regions on the ($|V_{e N}|^2, |V_{\mu N}|^2$) parameter space for the case of LNC decays mediated by Dirac neutrinos for $\mathcal{B}r^{\rm D}(\Lambda_b^0 \to p \mu \mu e \nu) < 10^{-10}, 10^{-11} $  and $\mathcal{B}r^{\rm D}(\Lambda_b^0 \to \Lambda_c  \mu \mu e \nu) < 10^{-9}, 10^{-10} $ for the fixed values of $\tau_{N_1} = \tau_{N_2} = [100,1000]$ ps, $\theta_{21}^{\rm LNV}= \pi/3$ , $\Delta M_N = 5\times 10^{-14} $ GeV, $\beta_N\gamma_N= 20$,  $L= 100$ cm, and $m_{N_1}= 2$ GeV and 3.2 GeV for $\Lambda_c$ and proton respectively.  Vertical and horizontal lines correspond to the present upper bounds on heavy-to-light mixing elements. \label{fig:cons2}}
\end{center}
\end{figure}

\section{Numerical analysis}\label{Numerical}
For numerical analysis, $\Lambda_b^0\to \Lambda_c^+$ and $\Lambda_b^0\to p^+$ form factors are taken from the lattice QCD calculations presented in \cite{Detmold:2015aaa}. We leave the decay widths of the sterile neutrinos as phenomenological parameters that can be measured in the experiment. Since the sterile neutrinos are almost degenerate, we will assume their lifetimes to be approximately equal, $\tau_{N_1}\approx\tau_{N_2} = \tau_N$. We present our analysis for $\tau_N  = [100, 1000]$ps \cite{Das:2021prm}. For nearly degenerate neutrinos, it is also natural to assume $|V_{\ell N_1}|\sim |V_{\ell N_2}| \equiv |V_{\ell N}|$. Present experimental bounds on the heavy-to-light mixing are $|V_{e N}|^2\leq 10^{-7}$ and $|V_{\mu N}|^2\leq 10^{-5}$ \cite{Abada:2017jjx, Atre:2009rg}. The neutrinos are expected to travel nearly at the speed of light and we assume that the velocity $\beta_N \geq 0.9$ corresponds to $\beta_N\gamma_N \geq 2$.

The observables of interest are the CP-averaged branching ratio and the CP-asymmetry. If sterile neutrinos involved are of Majorana types then the final states will include both LNV and LNC states and the CP-averaged branching ratio is
\begin{eqnarray}
\mathcal{B}r^{\rm M} &\approx &  \frac{1}{2}\Bigg(\mathcal{B}r_{\mathcal{B}_1}^{\rm M,LNC, osc}+\mathcal{B}r_{\mathcal{B}_1}^{\rm M,LNV, osc}+\mathcal{B}r_{\bar{\mathcal{B}}_1}^{\rm M,LNC, osc}+\mathcal{B}r_{\bar{\mathcal{B}}_1}^{\rm M,LNV, osc}  \Bigg)\, .
\end{eqnarray}
The branching ratios are obtained from decay rates by dividing with the total decay rates of $\mathcal{B}_1$ baryons. For simplicity, we have assumed that $\mathcal{B}_1$ and $\overline{\mathcal{B}}_1$ have the same decay rates and that is why we have used $\approx$ in the above equation. The CP-asymmetry observable for Majorana induced decay is
\begin{eqnarray}
\mathcal{A}_{CP}^{\rm M}&=& \frac{\Big(\Gamma_{\mathcal{B}_1}^{\rm M,LNC, osc}+\Gamma_{\mathcal{B}_1}^{\rm M,LNV, osc}\Big)-\Big(\Gamma_{\mathcal{\bar{B}}_1}^{\rm M,LNC, osc}+\Gamma_{\mathcal{\bar{B}}_1}^{\rm M,LNV, osc}\Big)}{\Big(\Gamma_{\mathcal{B}_1}^{\rm M,LNC, osc}+\Gamma_{\mathcal{B}_1}^{\rm M,LNV, osc}\Big)+\Big(\Gamma_{\mathcal{\bar{B}}_1}^{\rm M,LNC, osc}+\Gamma_{\mathcal{\bar{B}}_1}^{\rm M,LNV, osc}\Big)}\, .
\end{eqnarray}
When Dirac neutrino is involved then there are only LNC final states and CP-averaged branching ratio and the CP-asymmetry are defined as 
\begin{eqnarray}
	\mathcal{B}r^{\rm D} &\approx & \frac{1}{2}\Bigg(\mathcal{B}r_{\mathcal{B}_1}^{\rm D,LNC, osc}+\mathcal{B}r_{\bar{\mathcal{B}}_1}^{\rm D,LNC, osc} \Bigg),\,\,\ 
\mathcal{A}_{CP}^{\rm D} = \frac{\Big(\Gamma_{\mathcal{B}_1}^{\rm D,LNC, osc}-\Gamma_{\mathcal{\bar{B}}_1}^{\rm D,LNC, osc}\Big)}{\Big(\Gamma_{\mathcal{B}_1}^{\rm D,LNC, osc}+\Gamma_{\mathcal{\bar{B}}_1}^{\rm D,LNC, osc}\Big)}\,.
\end{eqnarray}
We first consider decays occurring through a Majorana neutrino in the intermediate state so that there are both LNV and LNC final states. In figure we \ref{fig:brmNtheta} we show the CP-averaged branching ratios for $\Lambda_b^0\to (\Lambda,p)\mu\mu e \nu$ as a function of neutrino mass $m_{N_1}$. For simplicity we chose $\theta^{\rm LNV}_{21} = \theta^{\rm LNC}_{21}$ though in general the CP-odd pases can be different. The plots are obtained for neutrino displaced vertex length $L=100$ cm, $\gamma_N\beta_N =2$, $\tau_{N} = 100$ ps, and the mass difference is $\Delta M_N = 5\times 10^{-14}$ GeV. Maximum branching ratios are obtained at $m_{N_1} = 2$ GeV for $\Lambda_b^0\to\Lambda_c\mu\mu e\nu$ and at $m_{N_1} = 3.2$ GeV for $\Lambda_b^0\to p \mu\mu e\nu$ modes. In figure \ref{fig:brmNbeta}, for $\tau_N =100, 1000$ ps we show the CP-averaged branching ratios for $\Lambda_b^0\to (\Lambda_c,p)\mu\mu e\nu$ for different values of $\gamma_N\beta_N$ and for fixed values of $\theta^{\rm LNV}_{21}$ and $\theta^{\rm LNC}_{21}$. From the figures we find the CP-averaged branching ratios of $\Lambda_b^0 \to \Lambda_c \mu \mu e \nu$ lie between $10^{-10} - 10^{-9} $, whereas branching ratios for $\Lambda_b^0 \to p \mu \mu e \nu$ lie between $10^{-11} - 10^{-10} $. The one order of magnitude suppression in the $\Lambda_b^0\to p \mu\mu e\nu$ mode with respect to the $\Lambda_b^0\to \Lambda \mu\mu e\nu$ is attributed to Cabibbo-Kobayashi-Maskawa matrix elements. We find that the branching ratios for $\Lambda_b^0 \to (\Lambda_c,p) e e \mu \nu$ are one orders of magnitude suppressed compared to $\Lambda_b^0 \to (\Lambda_c,p) \mu \mu e \nu$ but the qualitative features are same as seen in figures \ref{fig:brmNtheta} and \ref{fig:brmNbeta}. Though LNV and LNC both are present in case of Majorana induced decay, LNV modes dominate over the LNC modes in the case of $\Lambda_b^0 \to (\Lambda_c,p)  \mu \mu e \nu $ but for the decay mode $\Lambda_b^0 \to (\Lambda_c,p)  e e \mu \nu $, LNC decays dominate  over LNV. This can be explained as follows: In $\Lambda_b^0 \to (\Lambda_c,p)  \mu \mu e \nu $ decays the LNV modes are proportional to $|V_{\mu N}|^4 \sim 10^{-10}$ where as the LNC branching ratios are proportional to $|V_{\mu N}|^2 |V_{e N}|^2 \sim10^{-12}$. In $\Lambda_b^0 \to (\Lambda_c,p)ee\mu\nu$ decays the LNV modes are proportional to $|V_{e N}|^4\sim 10^{-14}$ and the LNC modes are proportional to $|V_{\mu N}|^2|V_{e N}|^2 \sim10^{-12}$.

To compare branching ratios mediated by Dirac neutrinos, in figure \ref{fig:brmNbetaDirac} we show the CP averaged branching ratios for $\Lambda_b^0\to (\Lambda,p)\mu\mu e \nu$ as a function of $m_{N_1}$. The plots are shown for $\tau_{N} =100$ ps where as the rest of the parameter choices are same as in figure \ref{fig:brmNbeta}. Similar plots are obtained for $\mathcal{B}r(\Lambda_b^0\to (\Lambda,p)ee\mu \nu)$ modes. Qualitatively, the plots are similar to what we obtained for these modes with Majorana neutrinos. For the Dirac case, we get the similar plot like \ref{fig:brmNtheta} but there are two order suppression of magnitude. Quantitatively, $\mathcal{B}r^{\rm D}(\Lambda_b^0 \to (\Lambda_c,p)  \mu \mu  e \nu)$ is two orders of magnitude smaller than $\mathcal{B}r^{\rm M}(\Lambda_b^0 \to (\Lambda_c,p)  \mu \mu  e \nu)$ where as $\mathcal{B}r^{\rm M}(\Lambda_b^0 \to (\Lambda_c,p) ee \mu \nu)$ and $\mathcal{B}r^{\rm D}(\Lambda_b^0 \to (\Lambda_c,p) ee \mu \nu)$ are of the same order of magnitude.

In figure \ref{fig:brL} we show the CP-averaged branching ratios as a function of maximal displaced vertex length $L$ for decays involving Majorana neutrinos. The Majorana neutrino mass $m_{N_1} =2$ GeV for $\Lambda_b^0\to \Lambda_c\mu\mu e\nu$ and $m_{N_1} =3.2$ GeV for $\Lambda_b^0\to p \mu\mu e\nu$ and the values of the rest of the parameters are shown in the legends of the plots. The troughs and crests are modulations resulting from the oscillation effect that is clearly visible from these plots.  These plots indicate that for sufficiently small detector length $L_{\rm det}<L$, a significant fraction of neutrinos will decay outside the detector leading to a small branching ratio. For $L_{\rm det}>L$, most neutrinos decay inside, and therefore the branching ratio saturates. From the figure \ref{fig:brL}, we also observe that saturation in $L$ can happen early for a small value of $\gamma_N \beta_N$. As we notice from the equation \eqref{eq:probability},  the acceptance factor is larger for the lower value of $\gamma_N\beta_N$.

In the left panel of figure \ref{fig:acpYthetea} we show CP-asymmetry observable $\mathcal{A}_{CP}$ for 
$\Lambda_b^0\to\Lambda_c \mu\mu e \nu$  as a function of $y$ for different values of $\theta^{\rm LNV}_{21}$ and $\theta^{\rm LNC}_{21}$ and $L$ = 100 cm. This plot demonstrates that the maximum value of CP-asymmetry is obtained when the CP-odd phases are $\pi/2$. From the left panel of figure \ref{fig:acpYthetea}, it is also seen that $\mathcal{A}_{CP}$ gradually decreases with increasing $y$ because the overlap between the two neutrinos decreases. In the right panel of figure \ref{fig:acpYthetea} we show $\mathcal{A}_{CP}$ with respect to $y$ for different values of maximal neutrino displaced vertex length $L$, CP-odd phases $\theta^{\rm LNV}_{21}$ = $\theta^{\rm LNC}_{21}$ = $\pi/2$, $\tau_N = 100$ps, and $\gamma_N\beta_N=20$. We observe that the maximum value of $\mathcal{A}_{CP}$ is obtained for $y<8$. A strong dependence of $\mathcal{A}_{CP}$ on maximum displaced vertex length $L$ is also observed. For large values of $L$ the modulation is suppressed as can be seen from \eqref{eq:finalLNV} and  \eqref{eq:finalLNC}.  In figure $\ref{fig:acpL}$, for the different values of $y$, we show the variation of $\mathcal{A}_{CP}$ with respect to $L$. The left and the right panels are for $\gamma_N\beta_N =20$ and $\gamma_N\beta_N =50$, respectively. This means that oscillations length is lower in the left plot. It is again observed that the maximum value of $\mathcal{A}_{\rm CP}$ is observed when maximal displaced vertex $L$ is comparable to $L_{\rm osc}$. For large $L$, the modulations are suppressed by the exponential term as can be seen from equations \eqref{eq:finalLNV} and  \eqref{eq:finalLNC}. Similar plots are also obtained for Majorana neutrino mediated $\Lambda_b^0 \to p \mu \mu e \nu$ and $\Lambda_b^0 \to (\Lambda_c, p) e e \mu  \nu$ decays. We reach similar conclusions when decays are due to Dirac neutrinos.

Even if the decay modes are not observed, upper limits on the branching ratios can be translated to constraints on the heavy-to-light mixing elements. For example, if upper limits on branching ratios  $\mathcal{B}r^{\rm M}(\Lambda_b^0 \to \Lambda_c \mu \mu e \nu)< 10^{-7}, 10^{-8}$ and $\mathcal{B}r^{\rm M}(\Lambda_b^0 \to p \mu \mu e \nu)< 10^{-8}, 10^{-9}$ is reported, then it can be translated to constraints on the $(m_{N_1}, |V_{\mu N}|^2)$ plane as shown in figure $\ref{fig:cons}$. To obtain this plot, we have neglected the LNC contributions for simplicity as it is two orders of magnitude suppressed. The light and dark shaded regions represent the exclusion regions correspond for $\tau_{N_1}=\tau_{N_2}$ = 100 ps and 1000 ps respectively. To compare our bounds with experimental results, we have superimposed in figure $\ref{fig:cons}$,
the exclusion limits coming from different experiments like LHCb \cite{Shuve:2016muy, Aaij:2014aba}, Belle \cite{Liventsev:2013zz}, L3 \cite{Adriani:1992pq}, Delphi \cite{Abreu:1996pa}, NA3 \cite{Badier:1985wg}, CHARM \cite{Vilain:1994vg}, NuTeV \cite{Vaitaitis:1999wq} and NA48 \cite{CERNNA48/2:2016tdo}.  In $\ref{fig:cons2}$ we show the constraints on the $(|V_{e N}|^2, |V_{\mu N}|^2)$ parameter space for $\mathcal{B}r^{\rm D}(\Lambda_b^0 \to \Lambda_c \mu \mu e \nu)< 10^{-9},10^{-10}$ and $\mathcal{B}r^{\rm D}(\Lambda_b^0 \to p \mu \mu e \nu)< 10^{-10},10^{-11}$. Here the mass of the Dirac neutrino $m_{N_1} = 2 $ GeV and 3.2 GeV for $\Lambda_c$ and $p$ respectively and the mass difference $\Delta M_N = 5 \times 10^{-14}$ GeV for both cases. The rest of the parameters are mentioned in the caption of the plots. The vertical and horizontal lines correspond to the present upper bounds on the heavy to light mixing elements.

At the LHC, about 5\% of total $b$-hadrons formed are $\Lambda_b^0$. The LHCb is already capable of studying rare decays of $\Lambda_b^0$ and has already observed $\Lambda_b^0 \to \Lambda \mu^+\mu^-$ \cite{Aaij:2015xza, Aaij:2018gwm}. The decay modes discussed in this paper may be within the reach of future LHCb. Even if the modes are not seen, the upper limits can still be translated to complementary bounds on the heavy-to-light mixing elements $|V_{\ell N}|^2$.
 
\section{Summary \label{sec:summary}}
In this paper we have studied $\Lambda_b^0 \to (\Lambda_c^+, p^+) \ell_1^-\ell_2^-\ell_3^+\nu$ and its conjugate decays mediated by two quasi-degenerate GeV-scale sterile neutrinos. Both Majorana and Dirac neutrinos have been considered and we have focused on kinematical regions where the neutrinos can be on-shell. The heavy neutrino acceptance factor and effects of oscillations are included in the decay rates. We find that the CP-averaged branching ratio for $\Lambda_b^0 \to (\Lambda_c, p) \ell_1\ell_2\ell_3\nu$ is one orders of magnitude suppressed compared to $\Lambda_b^0 \to (\Lambda_c, p) \ell_1\ell_2\ell_3\nu$ due to Cabbibo-Kobayashi-Maskawa suppression. For the $\Lambda_b^0 \to (\Lambda_c, p)\mu\mu e\nu$ mode, the Majorana induced transition is two orders of magnitude larger than the Dirac induced transition. For the $\Lambda_b^0 \to (\Lambda_c, p)ee \mu\nu$ mode, the branching ratios for Majorana and Dirac induced transitions are of the same order of magnitude. Numerically, we have explored the possibility of CP-violation and find that it is appreciable when the neutrino mass difference is of the order of their average decay widths.

\section*{Acknowledgements}
DD acknowledges the DST, Govt. of India for the INSPIRE Faculty Fellowship (grant number IFA16-PH170).  JD acknowledges the Council of Scientific and Industrial Research (CSIR), Govt. of India for SRF fellowship grant with File No. 09/045(1511)/2017-EMR-I. 

\appendix
\section{Five-body decays of $\mathcal{B}_1(p) (\rightarrow \mathcal{B}_2(k) \ell_1(p_1) \ell_2^-(p_2) \ell_3^+(p_3) \nu(p_\nu) $ )\label{sec:fivebody}}
The decays proceeds in two steps, $\mathcal{B}_1({b}) \to \mathcal{B}_2({c}/u)\ell_1^-N_j$ followed by the decay of the neutrinos $N_j\to \ell_2^-\ell_3^+\nu$. The direct $(D)$ or a crossed $(C)$ channel as shown in figure \ref{fig:feynman}. 
For $j^{\rm th}$ neutrino in the intermediate state, we can write the matrix elements for $D$ and $C$ channels as
\begin{align}
\mathcal{M}_{D_j}^{\rm LNV\pm} &= (G_F^2 V_{bq}^{(\ast)} m_{N_j}) \big( v_{X_i}^{\rm LNV,\pm}  \big) P_{D_j}^{\rm LNV} H_\nu^\pm L^{\nu\alpha \pm}_{\rm LNV, D}  l_{\rm X,\alpha}^{\rm LNV \pm} \nn\ \\
\mathcal{M}_{C_j}^{\rm LNV\pm} &= (G_F^2 V_{bq}^{(\ast)} m_{N_j}) \big( v_{X_i}^{\rm LNV,\pm}  \big) P_{C_j}^{\rm LNV} H_\nu^\pm L^{\alpha\nu\pm}_{\rm LNV, C}  l_{\rm X,\alpha}^{\rm LNV \pm}\nn\ \\
\mathcal{M}_{D_j}^{\rm LNC\pm} &= (G_F^2 V_{bq}^{(\ast)} m_{N_j} ) \big( v_{D_i}^{\rm LNC,\pm}  \big) P_{D_j}^{\rm LNC} H_\nu^\pm L^{\nu\alpha \pm}_{\rm LNC, D}  l_{\rm D,\alpha}^{\rm LNC \pm} \nn\ \\
\mathcal{M}_{C_j}^{\rm LNC\pm} &= (G_F^2 V_{bq}^{(\ast)} m_{N_j}) \big( v_{C_i}^{\rm LNC,\pm} \big) P_{C_j}^{\rm LNC} H_\nu^\pm L^{\alpha\nu\pm}_{\rm LNC, C}  l_{\rm C,\alpha}^{\rm LNC \pm} 
\end{align}
Here the superscripts `+' and `-' are for $\mathcal{B}_1(b) \rightarrow \mathcal{B}_2(q) \ell_1^-  \ell_2^-  \ell_3^+ \nu $ and its CP-conjugate mode $\bar{\mathcal{B}}_1(\bar{b}) \rightarrow \bar{\mathcal{B}}_2(\bar{q}) \ell_1^+  \ell_2^+ \ell_3^-\bar{\nu}$  respectively. The heavy to light mixing elements $v_{X_i}^{\rm Z,\pm} $ (Z=LNV or LNC) are given in following.
\begin{align}
v_{D_i}^{\rm LNC,+} &= V_{\ell_1 N_i}(V_{\ell_3 N_i})^\ast ,\,\,\,\ v_{C_i}^{\rm LNC,+}= V_{\ell_2 N_i}(V_{\ell_3 N_i})^\ast,\,\,\,\ v_{X_i}^{\rm LNC,-} = (v_{X_i}^{\rm LNC,+})^\ast \nn\ \\
v_{D_i}^{\rm LNV,+}&= v_{C_i}^{\rm LNV,+}=v_{X_i}^{\rm LNV,+}=V_{\ell_1 N_i}V_{\ell_2 N_i},\,\,\,\ v_{X_i}^{\rm LNV,-} = (v_{X_i}^{\rm LNV,+})^\ast ,\,\,\ X=C, D
\end{align}
The leptonic part of the amplitudes are
\begin{align}
& L^{\nu\alpha+}_{\rm LNV, D} = \bar{u}_{\ell_1}(p_1) \gamma^\nu \gamma^{\alpha}(1+\gamma_5) v_{\ell_2}(p_2)\, ,\quad L^{\nu\alpha-}_{\rm LNV, D} = \bar{v}_{\ell_2}(p_2) \gamma^{\alpha}\gamma^\nu(1-\gamma_5) u_{\ell_1}(p_1)\nn\ \\
& L^{\alpha\nu+}_{\rm LNV, C} = \bar{u}_{\ell_1}(p_1)  \gamma^{\alpha} \gamma^\nu(1+\gamma_5) v_{\ell_2}(p_2)\, ,\quad L^{\alpha\nu-}_{\rm LNV, C} = \bar{v}_{\ell_2}(p_2) \gamma^\nu\gamma^{\alpha}(1-\gamma_5) u_{\ell_1}(p_1)\nn\ \\
& L^{\nu\alpha+}_{\rm LNC, D} = \bar{u}_{\ell_1}(p_1) \gamma^\nu\slashed{k}_N \gamma^{\alpha}(1-\gamma_5) v_{\ell_3}(p_3)\, ,\quad L^{\nu\alpha-}_{\rm LNC, D} = \bar{v}_{\ell_3}(p_3)\gamma^{\alpha} \slashed{k}_N \gamma^\nu(1-\gamma_5) u_{\ell_1}(p_1)\nn\ \\
& L^{\alpha\nu+}_{\rm LNC, C} = \bar{u}_{\ell_1}(p_2) \gamma^\nu\slashed{k}_N \gamma^{\alpha}(1-\gamma_5) v_{\ell_3}(p_3)\, ,\quad L^{\alpha\nu-}_{\rm LNC, C} = \bar{v}_{\ell_3}(p_3)\gamma^{\alpha} \slashed{k}_N \gamma^\nu(1-\gamma_5) u_{\ell_1}(p_2)\, .\end{align}
Using the the lattice QCD parametrizations of the form factors given in \cite{Detmold:2015aaa} we get the expression of the hadronic amplitude as
\begin{align}
& H^\mu = \langle \mathB2(k,s_k)|\bar{c}\gamma^\mu(1-\gamma_5) b |\mathB1(p,s_p)\rangle \nn\\
&= \bar{u}(k,s_k)\Big(A_1 q^\mu+A_2 k^\mu+A_3\gamma^\mu+\gamma_5 \big\{A_4q^\mu+A_5 k^\mu+A_6\gamma^\mu\big\} \Big)u(p,s_p)\, .
\end{align}
where we have used $p=q+k$ we can even write where the $q^2$ dependent functions are given as
\begin{align}
& A_1 = f_t^V \frac{\mB1-\mB2}{q^2}+ f_0^V \frac{\mB1+\mB2}{s_+}  \bigg(1- \frac{\mmB1-\mmB2}{q^2} \bigg) - f_\perp^V \frac{2\mB2}{s_+} \, ,\nn\ \\
& A_2 = 2 f_0^V  \frac{\mB1+\mB2}{s_+} - f_\perp^V \bigg(\frac{2\mB2}{s_+}+\frac{2\mB1}{s_+}\bigg) \, ,\nn\ \\
& A_3=f_\perp^V\, ,\nn\ \\
& A_4 = f_t^A \frac{\mB1+\mB2}{q^2} + f_0^A \frac{\mB1-\mB2}{s_-}  \bigg(1- \frac{\mmB1-\mmB2}{q^2} \bigg) + f_\perp^A \frac{2\mB2}{s_-} \, ,\nn\ \\
& A_5 = 2 f_0^A  \frac{\mB1-\mB2}{s_-} + f_\perp^A \bigg(\frac{2\mB2}{s_-} - \frac{2\mB1}{s_-}\bigg) \, ,\nn\ \\
& A_6=f_\perp^A\, .
\end{align}
With reference to figure~\ref{fig:feynman} the decay of the $W^+$ and $W^-$ for LNV and LNC processes respectively proceeds through the following currents
\begin{align}
&l_{\rm X,\alpha}^{\rm LNV +} = \bar{u}_{\nu_{\ell_3}}\gamma_\alpha(1-\gamma_5)v_{\ell_3}   \, ,\quad 
l_{\rm X,\alpha}^{\rm LNV -} = \bar{u}_{\ell_3}\gamma_\alpha(1-\gamma_5)v_{\bar{\nu}_{\ell_3}}=(l_{\rm X,\alpha}^{\rm LNV,+})^{\dagger}\nn\ \\
&l_{\rm D, \alpha}^{\rm LNC +} = \bar{u}_{\ell_2}\gamma_\alpha(1-\gamma_5)v_{\bar{\nu}_{\ell_2}}   \, ,\quad 
l_{\rm D,\alpha}^{\rm LNC -} = \bar{u}_{\nu_{\ell_2}}\gamma_\alpha(1-\gamma_5)v_{\ell_2}=(l_{\rm D,\alpha}^{\rm LNC,+})^{\dagger}\nn\ \\
&l_{\rm C, \alpha}^{\rm LNC +} = \bar{u}_{\ell_1}\gamma_\alpha(1-\gamma_5)v_{\bar{\nu}_{\ell_1}}   \, ,\quad 
l_{\rm C, \alpha}^{\rm LNC -} = \bar{u}_{\nu_{\ell_1}}\gamma_\alpha(1-\gamma_5)v_{\ell_1}=(l_{\rm C,\alpha}^{\rm LNC, +})^{\dagger}.
\end{align}
and the functions $P_{{D(C)}_j}$ can be written as 
\begin{align}
P^{\rm LNV}_{D_j} &= \frac{1}{(k_N^2-m_{N_j}^2)+i\Gamma_{N_j} m_{N_j}},\,\, P^{\rm LNV}_{C_j} = \frac{1}{(k_N^{\prime 2}-m_{N_j}^2)+i\Gamma_{N_j} m_{N_j}},\, \,\\
 P^{\rm LNC}_{D_j} &= \frac{1}{m_{N_j}} P^{\rm LNV}_{D_j},\,\,\ P^{\rm LNC}_{C_j} = \frac{1}{m_{N_j}} P^{\rm LNV}_{C_j}.
\end{align}
where $k_N= q-p_1$ and $k_N^\prime=q-p_2$ and $q=p-k$.

The matrix element mod-squared of the total amplitude, averaged over the initial spin and summed over the final spin is given by 
\begin{align}
|\overline{\mathcal{M}}_{tot}^{\rm Z, \pm} |^2&=  \frac{1}{2}	\sum_{spins}\bigg|\mathcal{M}_{\rm tot}^{\rm Z,\pm} \bigg|^2\, \nn\\ 
&= \frac{1}{2}	\sum_{spins}\bigg|\sum_{j=1}^2\big(\mathcal{M}_{D_j}^{\rm Z, \pm} + \mathcal{M}_{C_j}^{\rm Z, \pm}\big) \bigg|^2\, ,\nn\\
& = \frac{1}{2}	\sum_{spins}\bigg[\sum_{i,j=1}^2 \mathcal{M}_{D_i}^{\rm Z, \pm} (\mathcal{M}_{D_j}^{\rm Z, \pm})^\ast + \sum_{i,j=1}^2 \mathcal{M}_{C_i}^{\rm Z, \pm} (\mathcal{M}_{C_j}^{\rm Z, \pm})^\ast + \sum_{i,j=1}^2 \mathcal{M}_{D_i}^{\rm Z, \pm} (\mathcal{M}_{C_j}^{\rm Z, \pm})^\ast + \sum_{i,j=1}^2 \mathcal{M}_{C_i}^{\rm Z, \pm} (\mathcal{M}_{D_j}^{\rm Z, \pm})^\ast \bigg]\, ,\label{eq:Mtot}\nn\ \\
& =N\bigg[ \sum_{i,j=1}^2 v_{D_i}^{\rm Z\pm} (v_{D_j}^{\rm Z\pm})^\ast m_{N_i}  m_{N_j} P_{D_i}^Z (P_{D_j}^Z)^\ast T_{\rm Z}^\pm(DD^\ast) +\sum_{i,j=1}^2 v_{D_i}^{\rm Z\pm} (v_{C_j}^{\rm Z\pm})^\ast m_{N_i}  m_{N_j} P_{D_i}^Z (P_{C_j}^Z)^\ast T_{\rm Z}^\pm (DC^\ast)\nn\ \\
&+ (D \leftrightarrow C )\bigg]\, ,
\end{align}
where the sum over $spins $ refers to the sum over spins of all external particles. The normalization constant is
\begin{equation}
N =\frac{1}{2}G_F^4 |V_{bq}|^2\, ,
\end{equation}
and $T^{\rm Z, \pm}(DD^\ast)^\pm $,  $T^{\rm Z, \pm}(CC^\ast) $,  $T^{\rm Z, \pm}(DC^\ast)$,  $T^{\rm Z, \pm}(D^\ast C) $ are given in following
\begin{align}
T_{\rm Z}^\pm(DD^\ast) & = \sum_{spins}[H^\pm_\nu(H^\pm_\rho)^\ast] \sum_{spins}[L_{\rm Z,D}^{\nu\alpha\pm}(L_{\rm Z,D}^{\rho\beta\pm})^\ast] \sum_{spins} [l_{\rm D,\alpha}^{\rm Z,\pm} (l_{\rm D,\beta}^{\rm Z,\pm})^\ast]\\
T_{\rm Z}^\pm(CC^\ast) & =\sum_{spins}[H^\pm_\nu(H^\pm_\rho)^\ast] \sum_{spins}[L_{\rm Z,C}^{\alpha\nu\pm}(L_{\rm Z,C}^{\beta\rho\pm})^\ast] \sum_{spins} [l_{\rm C,\alpha}^{\rm Z,\pm} (l_{\rm C,\beta}^{\rm Z,\pm})^\ast]\\
T_{\rm Z}^\pm(DC^\ast)&= \sum_{spins}[H^\pm_\nu(H^\pm_\rho)^\ast] \sum_{spins}[L_{\rm Z,D}^{\nu\alpha\pm}(L_{\rm Z,C}^{\beta\rho\pm})^\ast] \sum_{spins} [l_{\rm D,\alpha}^{\rm Z,\pm} (l_{\rm C,\beta}^{\rm Z,\pm})^\ast]\\
T_{\rm Z}^\pm(D^\ast C)&= \sum_{spins}[(H^\pm_\nu)^\ast H^\pm_\rho] \sum_{spins}[(L_{\rm Z,D}^{\nu\alpha\pm})^\ast L_{\rm Z,C}^{\beta\rho\pm}] \sum_{spins}[(l_{\rm D,\alpha}^{\rm Z,\pm})^\ast l_{\rm C,\beta}^{\rm Z,\pm}]
\end{align}
From these above equations we get the following equality. 
\begin{eqnarray}
 T_{\rm Z}(DD^\ast)\equiv T_{\rm Z}^+(DD^\ast)=T_{\rm Z}^-(DD^\ast),\,\ T_{\rm Z}(CC^\ast)\equiv T_{\rm Z}^+(CC^\ast)=T_{\rm Z}^-(CC^\ast)
\end{eqnarray}

The total decay rates are given in equations \eqref{eq:b1rate}, \eqref{eq:b1barrate}, \eqref{eq:b3rate}, and \eqref{eq:b4barrate}. With reference to the equations, the expressions of $\widehat{\Gamma}_{\rm Z}^\pm(XY^\ast)_{ij}$ are given as
\begin{align}
\widehat{\Gamma}_{\rm Z}(XX^\ast)_{ij}&=\frac{N}{2\mB1 }\int m_{N_i} m_{N_j} P^Z_{X_i}(P_{X_j}^Z)^\ast T_{\rm Z}(XX^\ast) d\Phi_5,\,\,\, X = C,D.\\
\widehat{\Gamma}_{\rm Z}^\pm(XY^\ast)_{ij}&=\frac{N}{2\mB1 }\int m_{N_i} m_{N_j} P^Z_{X_i}(P_{Y_j}^Z)^\ast T_{\rm Z}^\pm(XY^\ast) d\Phi_5,\,\,\, X \neq Y = C,D.
\end{align}
The five body phase space $d\Phi_5$ can be written as
\begin{equation}
d\Phi_5 \equiv d\Phi_5(\mathB1\to\mathB2\ell_1\ell_2\ell_3\nu) = d\Phi_3(\mathB1\to\mathB2\ell_1 N) \frac{dp_N^2}{2\pi} d\Phi_3(N\to \ell_2\ell_3\nu)\, .
\end{equation}
%
%

\section{Three body decay of $\mathcal{B}_1(p)(\to\mathcal{B}_2(k)\ell_1(p_1)N(p_N))$ }
With reference to the formula \eqref{eq:aux1} the kinematics and the decay rate of $\mathcal{B}_1\to \mathcal{B}_2\ell_1^-N$ in the following subsections.
\subsection{Kinematics for $\mathcal{B}_1(p)\to\mathcal{B}_2(k)\ell_1(p_1)N(p_N)$ \label{sec:kinematics}}
We do the kinematics in the $\mathB1(b)$ rest frame (RF). In the $\mathB1$-RF the $\mathB2(c/u)$ and the first $W^-$ boson with four momentum $q$ fly in opposite direction. We say that the $\mathB2$ travels in the $+\hat{z}$-direction so that the four momentum of the $\mathB2$ and $W^-$ are 
\begin{align}
& k^{\mathB1\rm-RF} \equiv (m_{\mathB1}-E^{\mathB1\rm-RF}_q, 0, 0, {\bf k}^{\mathB1\rm-RF})\, ,\\
& q^{\mathB1\rm-RF} \equiv (E^{\mathB1\rm-RF}_q, 0, 0, -{\bf k}^{\mathB1\rm-RF})\, ,
\end{align}
where the $q^0$ and the ${\bf k}^{\mathB1\rm-RF}$ are 
\begin{align}
& E^{\mathB1\rm-RF}_q = \frac{\mmB1 + q^2 - \mmB2}{2\mB1}\, ,\quad |{\bf k}^{\mathB1\rm-RF}| = \frac{\sqrt{\lambda(\mmB1,\mmB2,q^2)}}{2\mB1}\, .
\end{align}
In the first $W^-$-RF the $\ell_1(p_1)$ and the $N(p_N)$ will decay back to back. We introduce an angle $\theta_1$ made by $\ell_1$ w.r.to the $\mathB2$ {\emph i.e.,} the $+\hat{z}$ direction. Hence the four moment of the $\ell_1$ and the $N$ in the ${W^--{\rm RF}}$ are 
\begin{align}
& p_1^{W^--{\rm RF}} = (E_1^{W^--{\rm RF}}, |{\bf p}_1^{W^--{\rm RF}}|\sin\theta_1, 0, |{\bf p}_1^{W^--{\rm RF}}|\cos\theta_1)\, ,\\
& p_N^{W^--{\rm RF}} = (\sqrt{q^2}-E_1^{W^--{\rm RF}}, -|{\bf p}_1^{W^--{\rm RF}}|\sin\theta_1, 0, -|{\bf p}_1^{W^--{\rm RF}}|\cos\theta_1)\, ,
\end{align}
where $E_1^{W^--{\rm RF}}$ and ${\bf p}_1^{W^--{\rm RF}}$ are given as 
\begin{align}
E_1^{W^--{\rm RF}} = \frac{q^2+m_1^2-p_N^2}{2\sqrt{q^2}}\, ,\quad |{\bf p}_1^{W^--{\rm RF}}| = \frac{\sqrt{\lambda(q^2, m_1^2,p_N^2)}}{2\sqrt{q^2}}\, .
\end{align} 

We will now boost back these two momentum from $W^-{\rm-RF}$ to $\mathB1{\rm-RF}$ using the following Lorentz boost matrix 
\begin{equation}
\Lambda_{W^- \to \mathB1} = \begin{pmatrix}
\gamma_1 & -\gamma_1\beta_{1x} & -\gamma_1\beta_{1y} & -\gamma_1\beta_{1z}\\ 
-\gamma_1\beta_{1x} & 1+(\gamma_1-1)\frac{\beta^2_{1x}}{\vec{\beta}_1^2} &  (\gamma_1-1)\frac{\beta_{1x}\beta_{1y}}{\vec{\beta}_1^2} & (\gamma_1-1)\frac{\beta_{1x}\beta_{1z}}{\vec{\beta}_1^2}\\
-\gamma_1\beta_{1y} & (\gamma_1-1)\frac{\beta_{1x}\beta_{1y}}{\vec{\beta}_1^2} &  1+(\gamma_1-1)\frac{\beta^2_{1y}}{\vec{\beta}_1^2} & (\gamma_1-1)\frac{\beta_{1y}\beta_{1z}}{\vec{\beta}_1^2}\\
-\gamma_1\beta_{1z} & (\gamma_1-1)\frac{\beta_{1x}\beta_{1z}}{\vec{\beta}_1^2} &  (\gamma_1-1)\frac{\beta_{1y}\beta_{1z}}{\vec{\beta}_1^2} & 1+(\gamma_1-1)\frac{\beta^2_{1z}}{\vec{\beta}_1^2}\\
\end{pmatrix}
\end{equation}
where the velocity $\vec{\beta}_1$ is the velocity of the $W^-(q)$ as seen in the $\mathB1$-RF
\begin{align}
& \gamma_1 = \frac{1}{\sqrt{1-\vec{\beta}_1^2}}\,,\quad \beta_{1x} = 0\, ,\quad \beta_{1y} = 0\, , \quad \beta_{1z} =\frac{|{\bf k}^{\mathB1\rm-RF}|}{E^{\mathB1\rm-RF}_q}.
\end{align}
The function $\lambda(a,b,c)$ is defined by, $\lambda(a,b,c) \equiv a^2 +b^2 +c^2 - 2ab-2ac-2bc$. 
\subsection{Matrix element and phase space for $\mathcal{B}_1(p)\to\mathcal{B}_2(k)\ell_1(p_1)N(p_N)$ \label{sec:matrix}}
Matrix element of $\mathcal{B}_1(b)\to\mathcal{B}(c/u)_2\ell_1^- N$ is written as
\begin{eqnarray}
\mathcal{M}(\mathcal{B}_1\rightarrow \mathcal{B}_2 \ell_1^- N)= \frac{G_F}{\sqrt{2}}  V_{bq} V_{\ell_1 N} \bar{u}_{\ell_1}(p_1)\gamma^\mu(1-\gamma_5)v_N(p_N)H_\mu.
\end{eqnarray}
Therefore the matrix mod square with average over initial baryon spin is given as  
\begin{eqnarray}
|\overline{\mathcal{M}}|^2 &=& \frac{1}{2}\sum_{spins}  \mathcal{M} \mathcal{M}^\ast=\frac{1}{2}\frac{G_F^2}{2}|V_{bq}|^2 |V_{\ell_1 N}|^2 H_{\nu\rho}L^{\nu\rho}.
\end{eqnarray}
Where hadronic tensor current $H_{\nu\rho}$ and  leptonic tensor current $L_{\nu\rho}$ are given by
\begin{eqnarray}
H_{\nu\rho}=\sum_{spins}H_\nu H_\rho^\ast,\,\,\,\ L_{\nu\rho}=2\tr[\slashed{p}_1\gamma_\nu \slashed{p}_N\gamma_\rho(1-\gamma_5)].
\end{eqnarray}
Where the form of hadronic amplitude $H_\mu$ is given in \ref{sec:fivebody}. The differential decay width for $\mathcal{B}_1\to\mathcal{B}_2\ell_1^- N$ is given by
\begin{eqnarray}
	d\Gamma(\mathcal{B}_1\rightarrow \mathcal{B}_2 \ell_1^- N)=\frac{1}{2\mB1}d\Phi_3(\mathcal{B}_1\rightarrow \mathcal{B}_2 \ell_1 N)|\overline{\mathcal{M}}|^2.
\end{eqnarray}
Where the three-body phase space is given by 
\begin{eqnarray}
	d\Phi_3(\mathcal{B}_1\rightarrow \mathcal{B}_2 \ell_1 N)&=&\frac{d^3 k}{2E_{\mathcal{B}_2}(\vec{p}_{\mathcal{B}_2})(2\pi)^3} \frac{d^3p_1}{2E_1(\vec{p}_1)(2\pi)^3} \frac{d^3p_N}{2E_N(\vec{p}_N)(2\pi)^3} (2\pi)^4\delta^4(p-k-p_1-p_N)\nonumber\\\
	&=& d\Phi_2(\mathcal{B}_1(p)\rightarrow \mathcal{B}_2(k) W^-(q)) \frac{dq^2}{2\pi} d_2( W^-(q)\rightarrow \ell_1(p_1)N(p_N)).
\end{eqnarray}
Two particles phase space are given as follows.\\
\begin{eqnarray}
	d\Phi_2\big(\mathcal{B}_1(p)\rightarrow \mathcal{B}_2(k) W^-(q)\big)&=& \frac{1}{32\pi^2}\lambda^{1/2}\bigg(1,\frac{\mB2^2}{\mB1^2}, \frac{q^2}{\mB1^2} \bigg)d\Omega_{\hat{q}} \\
	d\Phi_2\big( W^-(q)\rightarrow \ell_1(p_1)N(p_N)\big) &=& \frac{1}{32\pi^2}\lambda^{1/2}\bigg(1,\frac{m_1^2}{q^2}, \frac{m_N^2}{q^2} \bigg) d\Omega_{\hat{p_1}}
\end{eqnarray}
Where $\hat{q}$ is the unit vector along the direction of $\vec{q}$ in the $\Lambda_b^0$ rest frame and $\hat{p_1}$ is the unit vector along the direction of $\vec{p_1}$ in the $W^- $ rest frame. It is straight forward to show that $d\Omega_{\hat{q}}$ = 4$\pi$ and $d\Omega_{\hat{p_1}}$ = 2$\pi d\cos \theta_1 $, and $\theta_1 $ is defined in the  subsection \ref{sec:kinematics}.
Here we get the following identities in the three body decay width between particle and anti-particle modes.
\begin{eqnarray}
\Gamma(\mathcal{B}_1\rightarrow \mathcal{B}_2 \ell_1^- N)={\Gamma}(\mathcal{\bar{B}}_1\rightarrow \mathcal{\bar{B}}_2 \ell_1^+ \bar{ N}),\,\,\ {\Gamma}(N\rightarrow \ell_2^- \ell_3^+ \nu)={\Gamma}(\bar{N}\rightarrow \ell_2^+ \ell_3^- \nu)
\end{eqnarray}

%

\end{document}